\documentclass[twocolumn,showpacs,prb,citeautoscript,amsmath,amssymb,floatfix,eqsecnum]{revtex4-2}
\usepackage{amsmath,amssymb,xcolor}
\usepackage{bm}
\usepackage{graphicx}
\usepackage{psfrag}
\usepackage{bbold}
\usepackage{bbm}
\newcommand{\bd}{\bm}
\newcommand{\kv}{{\bm{k}}}
\newcommand{\qv}{{\bm{q}}}
\begin{document}

\title{Functional renormalization group approach to phonon modified criticality: \\
anomalous dimension of strain and non-analytic corrections to Hooke's law}

\author{Max O. Hansen, Julia von Rothkirch, and Peter Kopietz}
  
\affiliation{Institut f\"{u}r Theoretische Physik, Universit\"{a}t
  Frankfurt,  Max-von-Laue Strasse 1, 60438 Frankfurt, Germany}

\date{May 27, 2025}

\begin{abstract}
We study the interplay between critical isotropic elasticity and classical Ising criticality using a functional renormalization group (FRG) approach which is implemented  such that the volume is fixed 
during the entire renormalization group flow. 
For dimensions slightly smaller than four 
we use a simple
truncation of the FRG flow equations to recover the fixed points of the constrained Ising model: the Gaussian fixed point G, the Ising fixed point I, the renormalized Ising fixed point R, and the spherical fixed point S. We show that the fixed points R and S are both characterized by a finite anomalous dimension $y_{\ast}<0$ of strain fluctuations, implying that the energy dispersion of longitudinal acoustic phonons exhibits a non-analytic momentum dependence proportional to $k^{1-y_{\ast}/2}$ for small momentum $k$.  We also derive and solve flow equations for the free energy at constant strain and compute stress-strain relations in the vicinity of the fixed points. As a result, we reaffirm that Ising criticality, controlled by the fixed point I, is preempted by a bulk instability. Beyond that, we find that the stress-strain relation at R and S remains linear to leading order (Hooke's law), as long as the interaction between strain and Ising fluctuations is sufficiently weak. However, the finite anomalous dimension of strain fluctuations $y_{\ast}$ gives rise to non-analytic corrections to Hooke's law.
\end{abstract}

\maketitle

\section{Introduction}

How do lattice vibrations affect critical phenomena in solids? For classical phase transitions, this problem has been addressed in pioneering works by Fisher \cite{Fisher68} and by Larkin and Pikin \cite{Larkin69}. Fisher used a thermodynamic analysis close to the critical point
to show that the coupling of critical order-parameter fluctuations to phonons can change the numerical values of critical exponents, a phenomenon called {\it{Fisher renormalization.}} 
Larkin and Pikin \cite{Larkin69} studied the question of whether phonons can change a continuous magnetic phase transition to a discontinuous one and found that this is possible
if the specific heat exponent $\alpha$ is positive so that the specific heat becomes large close to the critical point.
In the 1970s, this problem was studied using perturbative renormalization group (RG) methods~\cite{Aharony73,Bergman76}. Aharony found that for dimensions $D$ slightly below $D=4$ the Ising transition remains continuous even for positive $\alpha$ as long as
the phonon coupling is sufficiently weak.  Bergman and Halperin~\cite{Bergman76} analyzed an Ising model coupled to a compressible cubic lattice using perturbative RG to first order in $\varepsilon = 4 - D$. They carefully examined the influence of external conditions, distinguishing between constant pressure and constant volume, and implemented a
momentum shell RG procedure with periodic boundary conditions. This corresponds to keeping the volume fixed during the RG transformation.   
Apart from the Gaussian fixed point G and the Ising fixed point I, 
Bergman and Halperin \cite{Bergman76} found
two additional fixed point R and S, which are related to I and G through
Fisher renormalization of the corresponding critical exponents.
Moreover,
they found that Ising criticality controlled by the Ising fixed point I is preempted by a bulk instability, while 
the renormalized Ising fixed point R is characterized by a negative specific heat exponent $\alpha_R$, signaling the absence of a bulk instability. 
More recently, the coupling to lattice vibrations has been employed to describe deviations from Hooke's law near the critical point in the context of Mott metal-insulator transitions for a quasi-two-dimensional organic charge-transfer salt~\cite{Zacharias12}. The influence of lattice vibrations on quantum phase transitions has been explored in Refs.~[\onlinecite{Zacharias15,Zacharias15b,Chandra20,Samanta22,Sarkar23}].

In this work, we reexamine the effect of elasticity on classical Ising criticality using a functional renormalization group (FRG) approach \cite{Wetterich93,Berges02,Pawlowski07,Kopietz10,Dupuis21}. For simplicity, we assume that critical order-parameter fluctuations can be described by a classical scalar $\phi^4$-theory coupled to phonons via a singlet of the strain fluctuations. Employing a straightforward truncation of the FRG flow equations for the irreducible vertices of this coupled field theory, we recover the four fixed points G, I, R, and S of the constrained Ising model \cite{Rudnick74, Bergman76} mentioned above. By carefully deriving the proper field rescaling factor of the strain field representing elastic fluctuations, we show that R and S are characterized by a finite anomalous dimension
$y_{\ast} < 0$, already to leading order in $\varepsilon$. Although the fixed points R and S have been identified long ago using perturbative renormalization group methods to leading order in $\varepsilon$ \cite{Rudnick74, Bergman76}, the fact that these fixed points are characterized by a finite anomalous dimension $y_{\ast}<0$ has, to our knowledge, not been previously recognized. 
Consequently, at the critical points controlled by R and S, the energy dispersion of longitudinal phonons vanishes as $ k^{ 1 - y_{\ast} /2 }$ for small momenta $k$.
Moreover, our FRG approach includes a flow equation for the free energy at constant strain from which we
derive the elastic equation of state and the deviation of Hooke's law in the vicinity of the fixed points. In particular, we show that the non-analytic behavior of the elastic equation of state at R and S is determined by the
negative specific heat exponent associated with these fixed points.
Finally, by including various types of higher-order vertices which are not present in the bare action, we present evidence that the fixed points found to leading order in
$\varepsilon=4-D$ survive in three dimensions.

The rest of this work is  organized as follows: In Sec.~\ref{sec:Setup} we define our model and derive the effective action describing fluctuations of an Ising order parameter $\phi$ 
which is coupled to fluctuations of a strain singlet $E$. In Sec.~\ref{sec:FRG4} we derive 
a truncated system of FRG 
flow equations for the irreducible vertices of our model, retaining only those vertices which
are relevant at the fixed points to first order in $\varepsilon$. Although our flow equations are equivalent to those derived by Bergman and Halperin \cite{Bergman76}, our derivation using the established  FRG 
machinery is technically more transparent and clarifies the role of the field-renormalization
of the strain field. 
In Sec.~\ref{sec:fixed4} we discuss the fixed points of the RG flow for small
$\varepsilon = 4-D$. Using a more sophisticated truncation of the FRG flow equations,
in Sec.~\ref{sec:FRG3}
we estimate the fate of the fixed points in three dimensions.
In  Sec.~\ref{sec:free_energy} we use our FRG formalism to derive the 
scaling form of the free energy in the vicinity of the fixed points and discuss the
resulting elastic equation of state. Finally, in Sec.~\ref{sec:summary_and_discussion} we summarize our main results and present our conclusions.
In two appendices, we give additional technical details: In Appendix A, we derive the
effective action of the strain singlet by carrying out a constrained functional integration over all elastic fluctuations. In Appendix B, we present a more advanced truncation of the FRG flow equations for our model, retaining all three-point and four-point vertices allowed by symmetry.

\section{Field theory for critical Ising fluctuations coupled to strain}
\label{sec:Setup}

Following Ref.~[\onlinecite{Zacharias15}] we consider a generic classical model for an Ising order parameter $\phi ( \bd{r})$ 
which is coupled to elastic fluctuations described by a strain field 
$E_{ij} ( \bd{r} )$.
The classical order parameter fluctuations in the absence of elastic fluctuations are described by the Ginzburg-Landau-Wilson action \cite{Ma76}
 \begin{equation}
 \hspace{-1mm} S_{\phi} = \int d^D r \left[  \frac{c_0}{2} ( {\mathbf{\nabla}} \phi(\bd{r}))^2 + 
 \frac{r_0}{2}   \phi^2(\bd{r}) +
 \frac{ u_0 }{ 4!} \phi^4(\bd{r}) \right].
 \label{eq:Sphi}
 \end{equation}
The elastic properties of the crystal are determined
by the action of the strain field tensor $E_{ij} ( \bd{r} )$, which  in harmonic approximation
has the general form 
 \begin{equation}
 S_{E} = \frac{1}{2}
 \int d^D r \sum_{ijkl} \left[  E_{ ij } ( \bd{r} ) 
 C_0^{ ij, kl } E_{kl} ( \bd{r} ) \right],
 \label{eq:SE}
 \end{equation}
where
$C_0^{ ij, kl}$ are the bare elastic constants. Following Larkin and Pikin \cite{Larkin69}, we decompose the strain field tensor $E_{ij} ( \bd{r} )$ into a
homogeneous macroscopic part $e_{ij}$ and a part  $\epsilon_{ ij} ( \bd{r} )$ carrying finite momentum,
 \begin{equation}\label{eq:Seperation}
 E_{ij } ( \bd{r} ) = e_{ ij } + \epsilon_{ij} ( \bd{r} ),
 \end{equation}
where the finite momentum part $\epsilon_{ij} ( \bd{r} )$ satisfies
\begin{equation}
 \int d^D r \epsilon_{ij} ( \bd{r} ) =0,
 \label{eq:avzero}
 \end{equation}
and
can be expressed 
in terms of the  derivatives of the components $u_i ( \bd{r} )$ 
of the displacement field $\bd{u} ( \bd{r} )$ via
 \begin{equation}\label{eq:relation_e_und_u}
  \epsilon_{ij} ( \bd{r} ) = 
 \frac{1}{2} \left[ \partial_{i} u_{j} ( \bd{r} ) +
 \partial_{j} u_{i} ( \bd{r} ) \right].
 \end{equation}
Expressing the elastic action $S_{E}$ in terms of
the macroscopic part $e_{ij}$ and the phonon part $\epsilon_{ij} ( \bd{r} )$
we obtain
\begin{equation} \label{eq:Sel2}
 S_{{E}}  = 
  \frac{\cal{V}}{2} \sum_{ijkl} e_{ij} C_0^{ ij , kl} e_{ kl}+  \frac{1}{2} \int d^D r   \sum_{ijkl} \epsilon_{ij} ( \bd{r} )     C_0^{ij,kl}
 \epsilon_{kl} ( \bd{r} ),
\end{equation}
where ${\cal{V}}$ is the volume of the system and 
we have used Eq.~(\ref{eq:avzero}). 
Following previous work~[\onlinecite{Zacharias12,Zacharias15,Zacharias15b}], we assume that Ising fluctuations couple to strain only via a certain linear combination
 \begin{equation}
E  ( \bd{r} ) =  \sum_{ij} \gamma^{ij} E_{ij} ( \bd{r} ),
 \label{eq:Eee}
 \end{equation}
where the dimensionless constants  $\gamma^{ij}$ are determined by the symmetry group of the crystal.
To linear order in the strain
the interaction between the Ising field and the elastic degrees of freedom is then of the form~\cite{Zacharias15}
 \begin{equation}\label{eq:allgemeines_3point_coupling}
 S_{\phi, E}
 =  \frac{g_0}{2} \int d^D r  
  \phi^2 (\bd{r} )  E ( \bd{r} ) , 
 \end{equation}
with bare coupling $g_0$.
In principle, we could also add to $S_{\phi, E }$ a  
hybridization term proportional to $\phi ( \bd{r} ) E ( \bd{r} )$.
While such a term is allowed in special cases, such as end-points of lines of first-order Mott transitions \cite{Zacharias15,Zacharias12}, we restrict our calculations to the case of a cubic coupling (\ref{eq:allgemeines_3point_coupling}) in this work.  
Moreover, we assume elastic isotropy 
for simplicity, which implies $\gamma^{ij}=\delta^{ij}$. The relevant combination of strain fields
\begin{equation}
	E(\mathbf{r})= \sum_{i} [ e_{ii}+\epsilon_{ ii}(\bd{r}) ] \equiv e + \epsilon ( \bd{r} ) 
\end{equation}
can then be identified with the volume (or bulk) strain.
Recall \cite{Landau91,Chaikin95} that for an isotropic system, the tensor of elastic constants can be parametrized in terms of two Lamé parameters $\lambda$ and $\mu$ as follows 
\begin{equation}\label{eq:Cijk_Lame}
	C_{0}^{ij;kl}=\lambda \delta^{ij}\delta^{kl}+\mu\left(\delta_{ik}\delta_{jl}+\delta_{il}\delta_{jk}\right),
\end{equation}
where $\mu$  denotes the shear modulus and relates to the bulk modulus $K_0$ as follows \cite{footnoteel},
\begin{equation}\label{eq:bulk_modulous}
	K_0=\lambda+\frac{2}{3}\mu.
\end{equation} 
At this point, it is convenient to work with the effective action of the volume strain field
$E ( \bd{r} )$ which can be obtained by performing a constrained functional 
integration over all strain fluctuations at fixed $E$. 
Technical details are given in Appendix A. Transforming the decomposition (\ref{eq:Eee}) to
momentum space, 
\begin{equation}
	E_{\kv}=   \int d^D r e^{ - i \bd{k} \cdot \bd{r} } E ( \bd{r} ) = \mathcal{V}e\delta_{\kv,0}+\left(1-\delta_{\kv,0}\right)\epsilon_{\kv},
\end{equation}
we obtain for the quadratic part of the effective action of the volume strain 
 \begin{align}
	S_{E}^{\rm eff}  & =\mathcal{V}\frac{K_0}{2}e^2+\frac{\rho_{0}}{2}\int_{\kv\neq0}\epsilon_{-\kv}\epsilon_{\kv}
\nonumber
\\
 &  = \frac{1}{2} \int_{\bd{k}} F_0^{-1} ( \bd{k}  ) E_{ - \bd{k} } E_{\bd{k}},	
\end{align} 
where 
$\int_{\bd{k}} = {\cal{V}}^{-1} \sum_{\bd{k}}$, the stiffness
\begin{equation}
	\rho_{0}=K_0+\frac{4}{3}\mu=\lambda+2\mu
	 \label{eq:rhodef}
\end{equation}
is proportional to the square of the longitudinal phonon velocity \cite{footnoteel},
and we have introduced the inverse volume strain propagator
 \begin{align}
F_0^{-1}(\kv)=&\delta_{\kv,0}K_0+(1-\delta_{\kv,0})\rho_{0}.\label{eq:free_E_propagator} 
 \end{align}
Transforming also the Ising field to momentum space, we obtain the following effective action
for Ising fluctuations coupled to volume strain, 
\begin{align}
	S[\phi,E] & =\frac{1}{2}\int_{\kv} 
   \left[ 	
	( G_0^{-1}(\kv) + r_0)  \phi_{-\kv}\phi_{\kv}+ F_0^{-1}(\kv)E_{-\kv}E_{\kv} \right]
	\nonumber 
	\\
	& +\frac{g_0}{2}\int_{\kv,\qv}^{}\phi_{-\kv}\phi_{\kv-\qv}E_{\qv} \nonumber \\
	 & +\frac{u_0}{4!}\int_{\kv,\kv',\qv}^{}\phi_{-\kv}\phi_{\kv-\qv}\phi_{-\kv'}\phi_{\kv'+\qv},\label{eq:Sbare}
\end{align}
where
	\begin{align}
		G_{0}^{-1}(\kv)=& c_0 k^2 .  \label{eq:free_phi_propagator}
	\end{align}
At this point, the reader might wonder why we do not include the bare mass $r_0$ into the definition of $G_0^{-1} ( \bd{k} )$. The reason is rather subtle and is related to a simplification of the RG flow of the free energy, as will be explained
at the end of Sec.~\ref{sec:FRG4}. Of course, the bare mass $r_0$ is not ignored but is taken into account via the initial condition of the
self-energy $\Sigma_{\Lambda} ( \bd{k} )$ of the Ising field,
see Eq.~(\ref{eq:selfinit}) below.
The action (\ref{eq:Sbare}) is  the starting point of our FRG calculations
in Sec.~\ref{sec:FRG4}.

In general, the effective action of the volume strain also includes anharmonic terms. Hence, the effective action ~(\ref{eq:Sel2}) should be extended to
\begin{align}
	S^{\rm eff}_E & = 
	\frac{1}{2} \int_{\bd{k}} F_0^{-1} ( \bd{k}  ) E_{ - \bd{k} } E_{\bd{k}}
	\nonumber
	\\
	& + \int d^Dr\left[\frac{ h_0 }{3!} E^3 ( \bd{r}) + \frac{ v_0}{ 4!} E^4 ( \bd{r} )  
	+ \ldots  \right],
	\label{eq:Sanharm}
\end{align}
where the ellipsis denotes terms of order $E^5$ and higher.
%
Moreover, when including strain fluctuations up to fourth order in the fields, we should also add  a mixed quartic interaction
$\phi^2 E^2$ 
to the action $S_{\phi , E}$ in Eq.~(\ref{eq:allgemeines_3point_coupling}),
\begin{equation}
	\hspace{-0mm} S_{\phi, E}=\int d^D r \left[\frac{g_0}{2}\phi^2 (\bd{r} ) E ( \bd{r} ) +\frac{w_0}{(2!)^2}\phi^2 (\bd{r} ) E^2 ( \bd{r} ) \right].
	\label{eq:Shigh}
\end{equation} 
In fact, even if the bare action does not include the couplings  $h_0$, $v_0$, and $w_0$, 
the RG flow generates interactions of this type.
Fortunately, for small $\varepsilon = 4 - D$ these couplings are irrelevant 
at the four fixed points introduced above, so that in Sec.~\ref{sec:FRG4} and Sec.~\ref{sec:fixed4}  we will simply ignore
the corresponding interaction vertices.
However, in $D =3$ the effect of the
cubic coupling $h_0$ and the quartic couplings $v_0$ and $w_0$ on the RG flow is a priori not clear. Therefore, in Sec.~\ref{sec:FRG3} we will explicitly include those couplings in our FRG calculations, though we relegate all technical details in Appendix B.
Our final result is that the above higher-order interactions 
do not qualitatively modify the
fixed-point structure obtained to leading order in $\varepsilon = 4 -D$, further justifying that  we neglect the couplings $h_0$, $v_0$, and $w_0$ introduced 
in Eqs.~(\ref{eq:Sanharm}) and (\ref{eq:Shigh}) in the following two sections.

\section{Functional renormalization group flow equations}
\label{sec:FRG4}

\subsection{Exact Wetterich equation}

In this section, we derive exact FRG flow equations for the irreducible vertices 
of the classical field theory with action $S [ \phi, E ]$ given in (\ref{eq:Sbare}). Therefore, 
we add scale dependent regulators $R^{\phi}_{\Lambda} ( \bd{k} )$ and
$R^{E}_{\Lambda} ( \bd{k} )$
to the inverse propagators, which amounts to the substitution
 \begin{subequations}
 \label{eq:substitution}
 \begin{eqnarray}
 G_0^{-1} ( \bd{k} ) & \rightarrow &   G_{0,\Lambda}^{-1} ( \bd{k} ) =   G_0^{-1} ( \bd{k} ) + R^{\phi}_{\Lambda} ( \bd{k} ),
 \\
 F_0^{-1} ( {\bd{k}} ) & \rightarrow & F_{ 0 , \Lambda}^{-1} ( \bd{k} ) =  
F_0^{-1} ( {\bd{k}} ) + 
R^{E}_{\Lambda} ( \bd{k} ),
 \end{eqnarray}
 \end{subequations}
where $\Lambda$ is some momentum scale separating short-wavelength from long-wavelength fluctuations.
As usual, we require that for $\Lambda \rightarrow 0$ both  regulators $R^{\phi}_{\Lambda} ( \bd{k} )$
and  $R^{E}_{\Lambda} ( \bd{k} )$
vanish so that
we recover our original model, while for $\Lambda \rightarrow \infty$ the regulators diverge so that fluctuations 
on all scales are suppressed. Following the usual procedure \cite{Kopietz10} we now derive
formally exact flow equations for the scale-dependent irreducible vertices of our model. Therefore we first introduce the generating functional ${\cal{G}}_{\Lambda} [ J, \sigma]$ of the connected correlation functions of our model,
\begin{eqnarray}
	e^{ {\cal{G}}_{\Lambda} [ J , \sigma ] } = 
	\int {\cal{D}}[  \phi , E] ^{ - S_{\Lambda} [ \phi, \epsilon ] + 
		\int_{\bd{k}} [ J_{ - \bd{k}} \phi_{\bd{k}} +  \sigma_{ - \bd{k}} E_{\bd{k}} ] }.
\end{eqnarray} 
A functional Taylor expansion in powers of the sources $J ( \bd{r} )$ and
$\sigma ( \bd{r} )$ generates the correlation functions of the conjugate fields
$\phi ( \bd{r})$ and $E ( \bd{r} )$. Physically, the source $J ( \bd{r} )$ represents an inhomogeneous external magnetic field while the
source $\sigma ( \bd{r} )$ represents the an external stress conjugate to the volume strain $E ( \bd{r} )$.
The field expectation values for finite sources, therefore, are 
 \begin{subequations}
  \label{eq:fieldsource}
	\begin{eqnarray}
		\bar{\phi}_{\bd{k}} &=& \langle \phi_{ \bd{k}}\rangle = \frac{ \delta {\cal{G}}_{\Lambda} [ J , \sigma ] }{ \delta J_{ - \bd{k} } },
		\\
		\bar{E}_{\bd{k}} &=& \langle E_{ \bd{k}}\rangle = \frac{ \delta {\cal{G}}_{\Lambda} [J , \sigma ] }{ 
			\delta \sigma_{ - \bd{k} } }.
	\end{eqnarray}
\end{subequations}
The generating functional of the scale-dependent irreducible vertices, also called the average affective action,  is in turn defined via the following subtracted Legendre transform,
 \begin{eqnarray}
 \Gamma_{\Lambda } [ \bar{\phi} , \bar{E} ] & = &  
 \int_{\bd{k}} \left[ J_{ - \bd{k}} \bar{\phi}_{\bd{k}}
 +  \sigma_{ - \bd{k}} \bar{E}_{\bd{k}} \right]
 - {\cal{G}}_{\Lambda} [ J , \sigma ]
 \nonumber
 \\
 & - & \frac{1}{2} \int_{\bd{k}} \left[ R^{\phi}_{\Lambda} ( \bd{k} ) \bar{\phi}_{ - \bd{k}} \bar{\phi}_{\bd{k}} 
 +  R^{E}_{\Lambda} ( \bd{k} ) \bar{E}_{ - \bd{k}} 
  \bar{E}_{\bd{k}} \right], 
 \hspace{8mm}
 \label{eq:Gammadef}
  \end{eqnarray}
where on the right-hand side we should express the sources $J$ and 
$\sigma$ as functionals of the field averages $\bar{\phi}$ and $\bar{E }$
by inverting the relations (\ref{eq:fieldsource}).

To obtain RG flow equations for fixed volume, we keep the expectation value of the volume strain $\bar{E}_{\bd{k} =0}$ constant during the RG flow, necessitating that the conjugate stress $\sigma_{\Lambda}$ depends on the RG scale $\Lambda$. A similar procedure has been adopted in Ref.~[\onlinecite{Chichutek21}]. 
To implement this formally, we set
 \begin{equation}
 \bar{E}_{\bd{k}} = \delta_{\bd{k},0} {E}_0 + \bar{\epsilon}_{\bd{k}} 
 = {\cal{V}} \delta_{\bd{k} , 0 } {e}_0  + \bar{\epsilon}_{\bd{k}} ,
 \end{equation}
where $E_0 = {\cal{V}} e_0$ is the externally fixed value of the homogeneous strain. The shifted generating functional
 \begin{equation}
 \Gamma_{\Lambda} [ \bar{\phi} , \bar{\epsilon} ; E_0 ] =
 \Gamma_{\Lambda} [ \bar{\phi} , \bar{E} = E_0 + \bar{\epsilon} ]
 \end{equation}
then satisfies the usual Wetterich equation \cite{Wetterich93}
 \begin{equation}
\hspace{-0mm}\partial_{\Lambda} {\Gamma}_{\Lambda} [ \bar{\phi} , \bar{\epsilon} ; E_0] = 
 \frac{1}{2} {\rm Tr} \left[ ( \mathbf{{\Gamma}}^{\prime \prime}_{\Lambda} [ \bar{\phi} , 
\bar{\epsilon} ; E_0 ] + \mathbf{R}_{\Lambda} )^{-1} \partial_\Lambda \mathbf{R}_{\Lambda} \right],
 \end{equation}
where the trace is over momentum space and the two field types, and
the bold symbols
$ \mathbf{{\Gamma}}^{\prime \prime}_{\Lambda} [ \bar{\phi} , \bar{\epsilon} ; E_0]$ and 
$\mathbf{R}_{\Lambda}$ are matrices in momentum space and field-type space 
with matrix elements given by
\begin{align}
 \bigl[  \mathbf{{\Gamma}}^{\prime \prime}_{\Lambda} [ \bar{\phi} , \bar{\epsilon} ; E_0 ] \bigr]_{ \bd{k} 
 \bd{k}^{\prime}}^{ a a^{\prime}}
  & =     \frac{ \delta {\Gamma}_{\Lambda} [ \bar{\phi} , \bar{\epsilon} ; E_0] }{ \delta \Phi^a_{\bd{k}}
 \delta \Phi^{ a^{\prime}}_{\bd{k}^{\prime}}},
 \label{eq:Wetterich}
 \end{align}
and
 \begin{align}
 \bigl[ \mathbf{R}_{\Lambda} \bigr]^{ a a^{\prime}}_{ \bd{k}  \bd{k}^{\prime}} &  =  
 {\cal{V}} \delta_{\bd{k} + \bd{k}^{\prime},0 } \delta^{ a a^{\prime}} \left[
 \delta^{a \phi} R_{\Lambda}^{\phi} ( \bd{k} )
 + \delta^{a \epsilon}  R_{\Lambda}^{E} ( \bd{k} ) \right].
 \end{align}
Here, the upper index $a = \phi , \epsilon$ labels the two field types with 
$\Phi^{a = \phi}_{\bd{k}} = \bar{\phi}_{\bd{k}} $ and
$ \Phi^{a = \epsilon}_{\bd{k}} =   \bar{\epsilon}_{\bd{k}}$.
By expanding both sides of the Wetterich equation (\ref{eq:Wetterich}) in powers of the fields, we obtain an infinite hierarchy of exact flow equations for the irreducible vertices of our model, which depend on the fixed macroscopic strain $E_0$.

\vspace{0mm}
\subsection{Exact and truncated flow equations}

For dimensions close to $D=4$  it is sufficient to approximate the average effective action $\Gamma_\Lambda [ \bar{\phi} , \bar{\epsilon} ; E_0 ]$ by the following truncated vertex expansion 
 \begin{widetext}
 \begin{eqnarray}
 \Gamma_{\Lambda} [ \bar{\phi} , \bar{\epsilon} ; E_0] & \approx  & \Gamma^{(0)}_{\Lambda}+\sigma_{\Lambda}\bar{\epsilon}_{\kv=0} + \frac{1}{2} \int_{\bd{k}}
 \left[  G_0^{-1} ( \bd{k} ) + \Sigma_{\Lambda} ( \bd{k} ) \right]
 \bar{\phi}_{ - \bd{k}} \bar{\phi}_{\bd{k}} 
 + \frac{1}{2} \int_{\bd{k}} \left[ F_0^{-1} ( {\bd{k}} ) + \Pi_{\Lambda} ( \bd{k} ) \right]
 \bar{\epsilon}_{ - \bd{k}} \bar{\epsilon}_{\bd{k}} 
 \nonumber
 \\
 & + &  \frac{1}{2} \int_{\bd{k}_1} \int_{\bd{k}_2} \int_{\bd{k}_3}
 {\cal{V}} \delta_{  \bd{k}_1 + \bd{k}_2 + \bd{k}_3 ,0}
 \Gamma_{\Lambda}^{\phi \phi \epsilon} ( \bd{k}_1 , \bd{k}_2 , \bd{k}_3 )
  \bar{\phi}_{\bd{k}_1 }  \bar{\phi}_{\bd{k}_2} \bar{\epsilon}_{\bd{k}_3}
 \nonumber
 \\
  & + &  \frac{1}{4!} \int_{\bd{k}_1}  \int_{\bd{k}_2}   \int_{\bd{k}_3}    \label{eq:free_E_propagator} \int_{\bd{k}_4} {\cal{V}}
 \delta_{ \bd{k}_1 + \bd{k}_2 + \bd{k}_3 + \bd{k}_4 ,0} \Gamma^{\phi \phi \phi \phi}_{\Lambda}
 ( \bd{k}_1 , \bd{k}_2 , \bd{k}_3 , \bd{k}_4 )
 \bar{\phi}_{\bd{k}_1 } \bar{\phi}_{\bd{k}_2} \bar{\phi}_{\bd{k}_3} \bar{\phi}_{\bd{k}_4} .
 \label{eq:vertexp}
 \end{eqnarray}
 \end{widetext}
To avoid a regulator-dependent contribution  $R_{\Lambda}^E ( \bd{k} =0 )$ $ \bar{\epsilon}_{\bd{k} =0}$ to the first functional derivative of
our generating functional $\Gamma_{\Lambda} [ \bar{\phi} , \bar{\epsilon} ; E_0 ]$ in Eq.~(\ref{eq:vertexp}) we
assume that only the finite-momentum components of the
strain field are regulated, which amounts to choosing the strain regulator in the form $R_{\Lambda}^E ( \bd{k} ) = ( 1 - \delta_{\bd{k} , 0 } ) R_{\Lambda}^E ( k / \Lambda )$. The reason why it is neither necessary nor convenient to regulate homogeneous fluctuations is that the contribution of homogeneous modes to the action is proportional to the
volume of the system, so that we may use the saddle point approximation to
obtain the corresponding contribution to the free energy \cite{Schuetz06}.

\begin{figure}[tb]
	\begin{center}
		\centering
		\vspace{0mm}
		\includegraphics[width=0.45\textwidth]{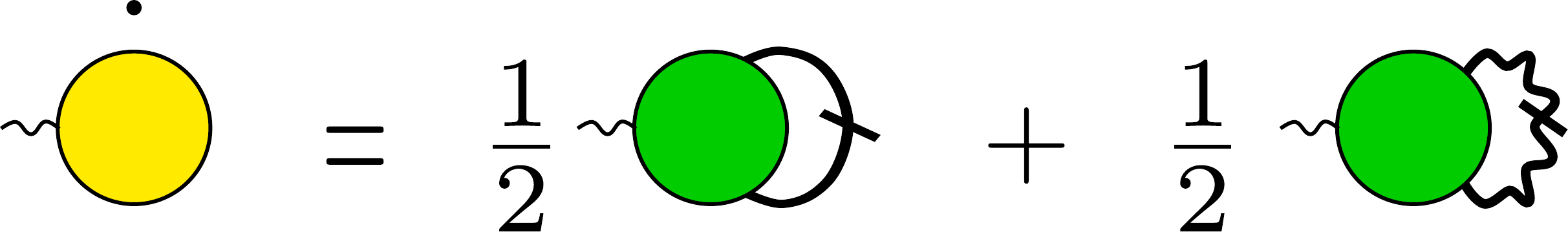}
	\end{center}
	  \vspace{-4mm}
	\caption{%
		Diagrammatic representation of the exact flow equation
(\ref{eq:stressflow})	
		for the 
		scale dependent stress $\sigma_{\Lambda}$, which is represented by a yellow circle with one wavy external leg. The dot above this circle represents the scale derivative $\partial_{\Lambda}$. On the right-hand side, the green circles represent the
		two types of three-point vertices $\Gamma^{\phi \phi \epsilon}_{\Lambda}$ and $\Gamma^{\epsilon \epsilon \epsilon }_{\Lambda}$ allowed by symmetry, where solid external legs
		represent the Ising field $\bar{\phi}$ while wavy external legs represent the strain field $\bar{\epsilon}$. Finally, the single-scale propagator $\dot{G}_{\Lambda}$ is represented by a slashed solid line, while $\dot{F}_{\Lambda}$ is represented by a slashed wavy line.		
		}
	\label{fig:flow_one_point}
\end{figure}

Exact flow equations for the vertices in Eq.~(\ref{eq:vertexp})
can now be obtained from the general hierarchy of flow equations 
given in Ref.~[\onlinecite{Kopietz10}].
The field-independent part  $\Gamma_{\Lambda}^{(0)}$ can be identified with the scale-dependent free energy (in units of temperature) at constant strain and satisfies the following exact flow equation
\begin{equation}\label{eq:free_energy_flow}
	\partial_{\Lambda}\Gamma_{\Lambda}^{(0)}=\frac{\mathcal{V}}{2}\int_{\kv}
	\left[  G_{\Lambda}(\kv)\partial_{\Lambda} R^{\phi}_{\Lambda}(\kv)+F_{\Lambda}(\kv)\partial_{\Lambda}R^{E}_{\Lambda}(\kv) \right],
\end{equation}
where  the regularized propagators $G_{\Lambda}(\kv)$ and $F_{\Lambda} ( \kv )$
are defined by
 \begin{subequations}
 \begin{align}
 G^{-1}_{\Lambda} ( \kv ) & = G^{-1}_{0, \Lambda} ( \kv ) + \Sigma_{\Lambda} ( \kv ) 
  \nonumber
  \\
  & 
 = G^{-1}_0 ( \kv ) + R_{\Lambda}^\phi ( \kv ) + \Sigma_{\Lambda} ( \kv )   ,
 \\
 F^{-1}_{\Lambda} ( \kv ) & = F^{-1}_{0, \Lambda} ( \kv ) + \Pi_{\Lambda} ( \kv ) 
  \nonumber
  \\
  & 
 = F^{-1}_0 ( \kv ) +    R_{\Lambda}^E ( \kv )  +  \Pi_{\Lambda} ( \kv ).
\end{align}
\end{subequations}
Below, we also need the corresponding single-scale propagators
 \begin{subequations}
 \begin{align}
 \dot{G}_{\Lambda} ( \bd{k} ) & = - G_{\Lambda}^2 ( \bd{k} ) \partial_{\Lambda}
 R_{\Lambda}^{\phi} ( \bd{k} ),
 \\
 \dot{F}_{\Lambda} ( \bd{k} ) & = - F_{\Lambda}^2 ( \bd{k} ) \partial_{\Lambda}
 R_{\Lambda}^{E} ( \bd{k} ).
 \end{align}
 \end{subequations}
The RG flow of the scale-dependent stress $\sigma_{\Lambda}$ is given by
\begin{align}
	\partial_\Lambda\sigma_{\Lambda}  =\frac{1}{2} \int_{\kv}^{}  \Bigl[
	&\dot{G}_{\Lambda}\left(\kv\right) {\Gamma}^{\phi \phi \epsilon}_{\Lambda}\left(-\kv,\kv,0\right)
	\nonumber
	\\
	 + 
	&\dot{F}_{\Lambda}\left(\kv\right) {\Gamma}^{\epsilon \epsilon \epsilon}_{\Lambda}\left(-\kv,\kv,0\right) \Bigr],
	\label{eq:stressflow}
\end{align}
which is shown diagrammatically in Fig.~\ref{fig:flow_one_point}.
Although the elastic three-point vertex
${\Gamma}^{\epsilon \epsilon \epsilon}_{\Lambda}\left(\kv_1,\kv_2, \kv_3 \right)$
is generated by the RG flow, we may neglect its contribution to leading order
in $4-D$ so that this vertex does not appear in the truncated vertex expansion
(\ref{eq:vertexp}).

\begin{figure}[tb]
 \begin{center}
  \centering
\vspace{2mm}
 \includegraphics[width=0.45\textwidth]{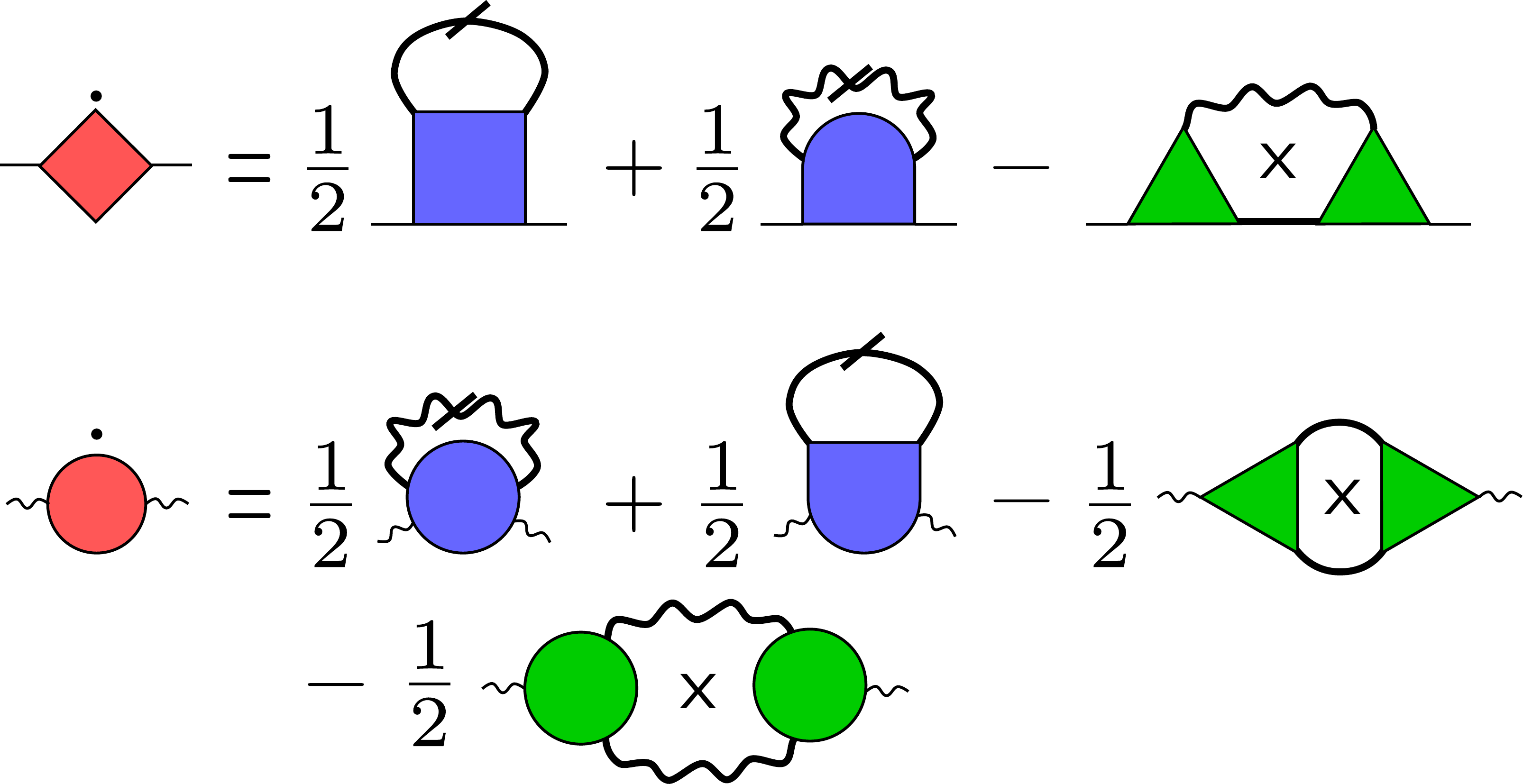}
   \end{center}
 \vspace{-4mm}
  \caption{%
Diagrammatic representation of the exact flow equations for the Ising self-energy $\Sigma_{\Lambda} 
( \bd{k} )$ (red square), and the strain self-energy  $\Pi_{\Lambda} ( \bd{k} )$ (red circle). 
The flow of the Ising self-energy is driven by
the Ising four-point vertex
 $\Gamma^{\phi \phi \phi \phi}_{\Lambda}$ (blue square),  the mixed four-point vertex $\Gamma_{\Lambda}^{\phi \phi \epsilon \epsilon}$ (blue half square/circle), and by the mixed three-point vertex
$\Gamma_{\Lambda}^{\phi\phi\epsilon}$ (green triangle). The flow of the elastic self-energy involves, in addition, the 
elastic three-point vertex
$\Gamma_\Lambda^{\epsilon \epsilon \epsilon}$ (green circle). The solid and wavy lines are explained in the caption of
Fig.~\ref{fig:flow_one_point}. The cross inside the loops means that
one should successively replace each of the lines forming the loop by
single-scale propagators using the product rule (\ref{eq:productrule}).
}
\label{fig:flow_two_point}
\end{figure}

The exact flow equations for the irreducible self-energy 
$\Sigma_{\Lambda} ( \kv )$ of the Ising field and the irreducible self-energy 
$\Pi_{\Lambda} ( \kv )$ of the strain field are shown diagrammatically in Fig.~\ref{fig:flow_two_point}. 
For small $4-D$, it is sufficient to retain only  the
vertices which appear in our truncated vertex expansion (\ref{eq:vertexp}), i.e., the mixed three-point vertex
$\Gamma^{\phi \phi \epsilon}_{\Lambda}$ and the
Ising four-point vertex $\Gamma^{\phi \phi \phi \phi}_{\Lambda}$.
In this approximation, the truncated flow equations for the self-energies are
\begin{align}
 \partial_{\Lambda} \Sigma_{\Lambda} ( \bd{k} ) & =  \frac{1}{2} \int_{\bd{q}} \dot{G}_{\Lambda}
 ( \bd{q} ) \Gamma_{\Lambda}^{ \phi \phi \phi \phi } ( - \bd{q} , \bd{q}, - \bd{k} , \bd{k} )
 \nonumber
 \\
 & - \int_{\bd{q}} [  {G}_{\Lambda} ( \bd{q} ) F_{\Lambda} ( \bd{q} + \bd{k} ) ]^{\bullet}
 \;
 \Gamma^{\phi \phi \epsilon}_{\Lambda} ( - \bd{q} , - \bd{k} , \bd{q} + \bd{k} )
 \nonumber
 \\
 & \hspace{12mm} \times 
   \Gamma^{\phi \phi \epsilon }_{\Lambda} (  \bd{q} ,  \bd{k} , - \bd{q} - \bd{k} ),
 \label{eq:Sigmaflow}
 \end{align}
\begin{align}
  \partial_{\Lambda} \Pi_{\Lambda} ( \bd{k} ) & = - \int_{\bd{q}} 
 \dot{G}_{\Lambda} ( \bd{q} ) G_{\Lambda} ( \bd{q} + \bd{k} )
  \Gamma_{\Lambda}^{\phi \phi \epsilon} ( \bd{q} + \bd{k} , - \bd{q} , - \bd{k} )
 \nonumber
 \\
 & \hspace{20mm} \times
   \Gamma_{\Lambda}^{\phi \phi \epsilon} ( -\bd{q} - \bd{k} ,  \bd{q} ,  \bd{k} ),
 \label{eq:Piflow}
 \end{align}
where  $ [ G F  ]^{\bullet}$ means that the terms in the bracket
should be successively promoted 
to a single scale propagator using the product rule
\begin{equation}
  [ {G}_{\Lambda} ( \kv) F_{\Lambda} ( \kv' ) ]^{\bullet} =
 \dot{G}_{\Lambda} ( \kv ) F_{\Lambda} ( \kv' )
 +   {G}_{\Lambda} ( \kv ) \dot{F}_{\Lambda} ( \kv' ) .
 \label{eq:productrule}
 \end{equation}

\begin{figure}[b]
	\begin{center}
		\centering
		\vspace{0mm}
		\includegraphics[width=0.45\textwidth]{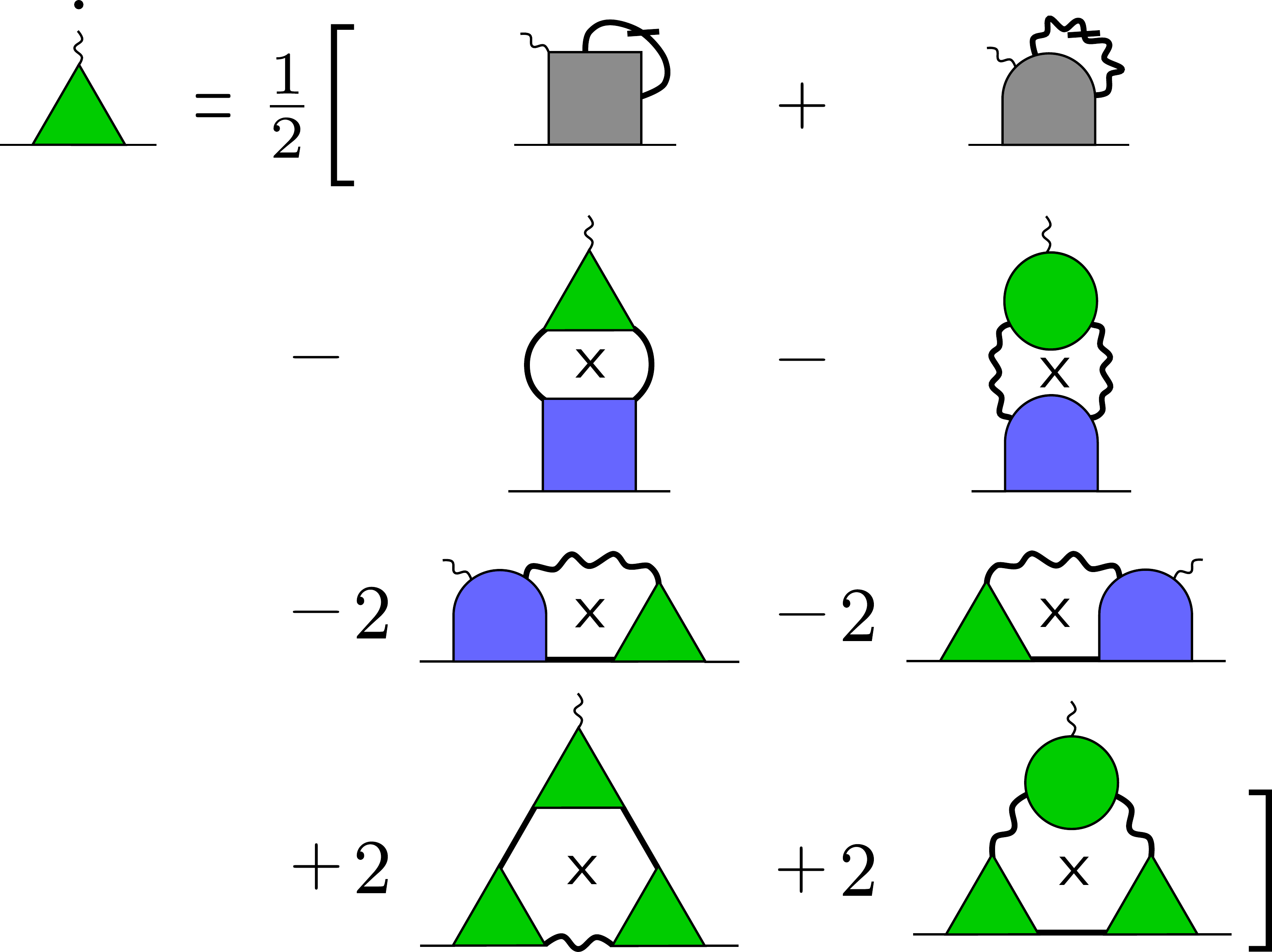}
	\end{center}
	\vspace{-4mm}
	\caption{%
		Diagrammatic representation of the exact FRG flow equation for the
		three-point vertex $\Gamma^{\phi \phi \epsilon}_{\Lambda}$. 
		The gray boxes represent two types of five-point vertices.
		The other
		symbols are defined in the captions of
		Figs.~\ref{fig:flow_one_point} and \ref{fig:flow_two_point}.
	}
	\label{fig:flow_three_point}
\end{figure}

\begin{figure}[t]
	\begin{center}
 	\centering
		\vspace{0mm}
		\includegraphics[width=0.45\textwidth]{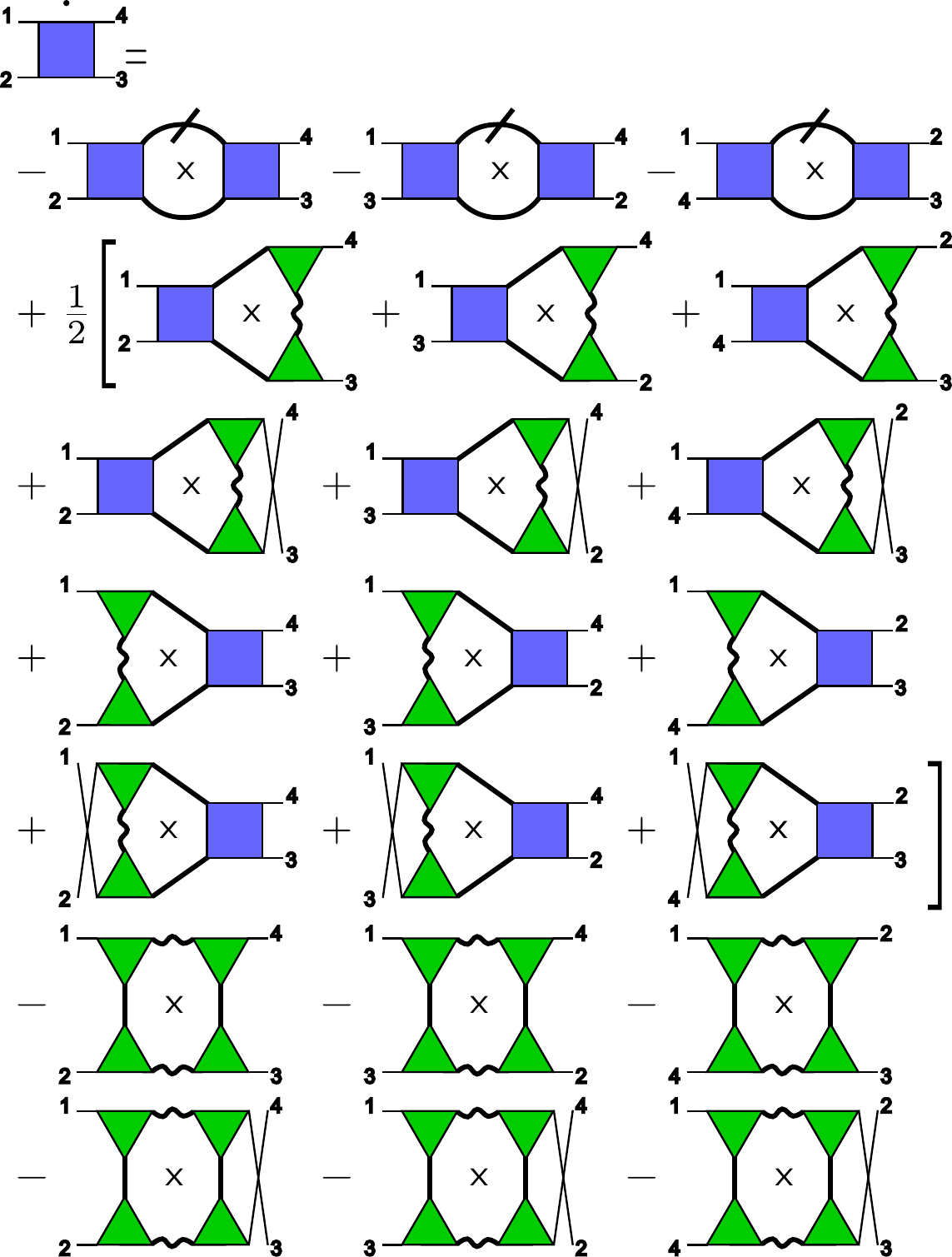}
	\end{center}
	  \vspace{-4mm}
	\caption{%
		Diagrammatic representation of the FRG flow equations for the
		Ising four-point vertex 
		$\Gamma^{\phi \phi \phi \phi}_{\Lambda} (
		\bd{k}_1 , \bd{k}_2 , \bd{k}_3 , \bd{k}_4 )$
		in a truncation where only the vertices that appear already in the
		bare action (\ref{eq:Sbare}) are retained.
	}
	\label{fig:flow_four_point}
\end{figure}

To obtain a closed system of flow equations within  the truncation (\ref{eq:vertexp}), we need  additional flow equations for the mixed three-point vertex 
$\Gamma^{\phi \phi \varepsilon}_{\Lambda}$ and the Ising four-point vertex
$\Gamma^{\phi \phi \phi \phi}_{\Lambda}$. The
exact flow equation for $\Gamma^{\phi \phi \varepsilon}_{\Lambda}$ is shown
diagrammatically in Fig.~\ref{fig:flow_three_point}.
Neglecting again the contribution from the vertices that do not appear in our truncation (\ref{eq:vertexp}), only two diagrams in Fig.~\ref{fig:flow_three_point} survive: The first one in the second line (one green triangle and blue square) and the first one in the last line (three green triangles). The resulting truncated flow equation for the mixed three-point vertex is
\begin{widetext}
 \begin{align}
 & \partial_{\Lambda} \Gamma^{\phi \phi \epsilon}_{\Lambda} ( \bd{k}_1 , \bd{k}_2 , \bd{k}_3 )
  = - \frac{1}{2} \int_{\bd{q}} [ {G}_{\Lambda} ( \bd{q} ) 
 G_{\Lambda} ( \bd{q} + \bd{k}_3 ) ]^{\bullet}
 \Gamma^{\phi \phi \phi \phi}_{\Lambda} ( \bd{k}_1 , \bd{k}_2 , \bd{k}_3 + \bd{q} , - \bd{q} )
 \Gamma^{\phi \phi \epsilon}_{\Lambda} ( \bd{q} , - \bd{q} - \bd{k}_3 , \bd{k}_3 )
 \nonumber
 \\
 &  + \int_{\bd{q}} \left[  {F}_{\Lambda} ( \bd{q} ) G_{\Lambda} ( \bd{q} + \bd{k}_1 ) 
 G_{\Lambda} ( \bd{q} - \bd{k}_2 ) \right]^{\bullet}  
\Gamma^{ \phi \phi \epsilon}_{\Lambda} ( \bd{k}_1 , \bd{q} , - \bd{q} - \bd{k}_1 )
 \Gamma^{ \phi \phi \epsilon}_{\Lambda} ( \bd{k}_2 , - \bd{q} ,  \bd{q} - \bd{k}_2 )
 \Gamma^{ \phi \phi \epsilon}_{\Lambda} ( \bd{q} + \bd{k}_1  , - \bd{q} + \bd{k}_2 ,  \bd{k}_3 ).
 \end{align}

Finally, we also need flow equation for the the properly symmetrized Ising four-point vertex $\Gamma^{\phi \phi \phi \phi}_{\Lambda} ( \kv_1 , \kv_2 , \kv_3, \kv_4) $. The exact flow equations involve vertices with up to six external legs. For our purpose, we only need the flow of
the Ising four-point vertex 
in the truncation, where only the vertices in 
Eq.~(\ref{eq:vertexp}) are retained.  
The relevant diagrams are shown in Fig.~\ref{fig:flow_four_point} and represent the following flow equation
 \begin{align}
  &  \partial_{\Lambda} \Gamma_{\Lambda}^{\phi \phi \phi \phi} ( \bd{k}_1 , \bd{k}_2 , \bd{k}_3 , \bd{k}_4 )
  = - \int_{\bd{q}}  \Bigl\{ \dot{G}_{\Lambda} ( \bd{q} ) G_{\Lambda} (\bd{q} + \bd{k}_1 + \bd{k}_2 )
 \Gamma_\Lambda^{\phi \phi \phi \phi} ( \bd{k}_1 , \bd{k}_2 , \bd{q} , - \bd{q} -  \bd{k}_1 - \bd{k}_2 )  
 \Gamma_\Lambda^{\phi \phi \phi \phi} ( - \bd{q} , \bd{q} - \bd{k}_3 - \bd{k}_4 , \bd{k}_3 , 
\bd{k}_4  )  
 \nonumber
 \\
 & \hspace{45mm}
+ ( \bd{k}_2 \leftrightarrow \bd{k}_3 ) + ( \bd{k}_2 \leftrightarrow \bd{k}_4 ) \Bigr\}
 \nonumber
 \\
& + \frac{1}{2} \int_{\bd{q}} \Bigl\{ \left[ G_{\Lambda} ( \bd{q} ) G_{\Lambda} ( \bd{q} + \bd{k}_1 
 + \bd{k}_2 ) G_{\Lambda} ( \bd{ q} + \bd{k}_1 + \bd{k}_2 + \bd{k}_3 ) \right]^{\bullet}
 \Gamma^{\phi \phi \phi \phi}_{\Lambda} ( \bd{k}_1 , \bd{k}_2 , \bd{q} , - \bd{q} - \bd{k}_1
 - \bd{k}_2 ) 
 \nonumber
 \\
 & \hspace{12mm} \times \Gamma^{\phi \phi \epsilon}_{\Lambda} ( \bd{q} + \bd{k}_1 + \bd{k}_2 , - \bd{q}
 - \bd{k}_1 - \bd{k}_2 - \bd{k}_3 , \bd{k}_3 ) \Gamma^{\phi \phi \epsilon}_{\Lambda}
 ( - \bd{q} , \bd{q} - \bd{k}_4 , \bd{k}_4 )
 + \mbox{($11$ permutations)} \Bigr\}
 \nonumber
 \\
 & -  \int_{\bd{q}} \Bigl\{ \left[ F_{\Lambda} ( \bd{q} ) 
 G_{\Lambda} ( \bd{q} + \bd{k}_1 ) F_{\Lambda} ( \bd{q} + \bd{k}_1 + \bd{k}_2 )
  G_{\Lambda} ( \bd{q} - \bd{k}_4 ) \right]^{\bullet}
 \Gamma^{\phi \phi \epsilon}_{\Lambda} ( \bd{k}_1 , - \bd{q} - \bd{k}_1 , \bd{q} )
 \Gamma^{\phi \phi \epsilon}_{\Lambda} ( \bd{k}_2 ,  \bd{q} +  \bd{k}_1 , - \bd{q} - \bd{k}_1 - \bd{k}_2 )
 \nonumber
 \\
 & \hspace{10mm}
 \times 
  \Gamma^{\phi \phi \epsilon}_{\Lambda} ( \bd{k}_3 ,  - \bd{q} + \bd{k}_4 , 
 \bd{q} +  \bd{k}_1 + \bd{k}_2 )
 \Gamma^{\phi \phi \epsilon}_{\Lambda} ( \bd{k}_4 , - \bd{q} , - \bd{k}_4 + \bd{q} )
 + \mbox{($5$ permutations)}
  \Bigr\}.
 \label{eq:flow4}
 \end{align}
 \end{widetext}
The permutations of the external momenta indicated in 
Eq.~(\ref{eq:flow4}) are explicitly drawn in 
Fig.~\ref{fig:flow_h_three_point} and render this vertex symmetric with respect to any permutation of the external momenta.

\subsection{Truncation for small $\varepsilon = 4-D$ with  sharp momentum cutoff}
\label{sec:sharp}

To construct the fixed points of the RG transformation to leading order
in $\varepsilon = 4-D$, it is sufficient to neglect the momentum-dependence of all vertices which amounts to approximating 
 \begin{equation}
 \Sigma_{\Lambda} ( \bd{k} )  \approx \Sigma_{\Lambda} (0), \; \; \; \; 
 \Pi_{\Lambda} ( \bd{k} )  \approx \Pi_{\Lambda} ( 0 ).
 \label{eq:SPinull}
 \end{equation}
Denoting the zero-momentum parts of the interaction vertices by
 \begin{subequations}
 \begin{eqnarray}
 g_{\Lambda} & = & \Gamma_{\Lambda}^{\phi \phi \epsilon} (0,0,0),
 \\
 u_{\Lambda} & = & \Gamma_{\Lambda}^{\phi \phi \phi \phi} ( 0,0,0,0),
 \end{eqnarray}
 \end{subequations}
our system of truncated flow equations reduces to
 \begin{subequations}
 \label{eq:flowtrunc}
 \begin{align}
 \partial_{\Lambda} \Gamma^{(0)}_{\Lambda} & = \frac{1}{2} \int_{\bd{q}}
  \left[
  G_{\Lambda} ( \bd{q} ) \partial_{\Lambda} R_{\Lambda}^{\phi} ( \bd{q} )
  +  F_{\Lambda} ( \bd{q} ) \partial_{\Lambda} R_{\Lambda}^{E} ( \bd{q} )
  \right],
\\ 
 \partial_{\Lambda} \sigma_{\Lambda}  & =\frac{g_\Lambda}{2} \int_{\bd{q}} \dot{G}_{\Lambda} ( \bd{q} ),
 \label{eq:stresssharp}
 \\
 \partial_{\Lambda} \Sigma_{\Lambda} (0) & =  \frac{u_{\Lambda}}{2} \int_{\bd{q}} \dot{G}_{\Lambda} ( \bd{q} )
 \nonumber
 \\
 & - g_{\Lambda}^2 \int_{\bd{q}} [ \dot{G}_{\Lambda} ( \bd{q} ) F_{\Lambda} ( \bd{q} ) +
 {G}_{\Lambda} ( \bd{q} ) \dot{F}_{\Lambda} ( \bd{q} ) ],
 \label{eq:Sigma0sharp}
 \\
 \partial_{\Lambda} \Pi_{\Lambda} (0) & = - g_{\Lambda}^2 \int_{\bd{q}} \dot{G}_{\Lambda} ( \bd{q} ) G_{\Lambda} ( \bd{q} ),
 \label{eq:Pi0sharp}
 \\
 \partial_{\Lambda} g_{\Lambda} & = - u_{\Lambda} g_{\Lambda} 
 \int_{\bd{q}} \dot{G}_{\Lambda} ( \bd{q} ) G_{\Lambda} ( \bd{q} )
 \nonumber
 \\ &
 + g_{\Lambda}^3 \int_{\bd{q}} [ 2 \dot{G}_{\Lambda} ( \bd{q} ) G_{\Lambda} ( \bd{q} ) 
 F_{\Lambda} ( \bd{q} ) +
  G_{\Lambda}^2 ( \bd{q} ) \dot{F}_{\Lambda} ( \bd{q} )  ],
 \\
 \partial_{\Lambda} u_{\Lambda} & = - 3 u_{\Lambda}^2  \int_{\bd{q}} \dot{G}_{\Lambda} ( \bd{q} ) G_{\Lambda} ( \bd{q} )
 \nonumber
 \\ & \hspace{-10mm}
 + 6 u_{\Lambda} g_{\Lambda}^2  \int_{\bd{q}} [ 2 \dot{G}_{\Lambda} ( \bd{q} ) G_{\Lambda} ( \bd{q} ) 
 F_{\Lambda} ( \bd{q} ) +
  G_{\Lambda}^2 ( \bd{q} ) \dot{F}_{\Lambda} ( \bd{q} )  ]
 \nonumber
 \\
& \hspace{-10mm} - 12 g_{\Lambda}^4  \int_{\bd{q}} [ \dot{G}_{\Lambda} ( \bd{q} ) G_{\Lambda} ( \bd{q} ) 
 F^2_{\Lambda} ( \bd{q} ) +
  G_{\Lambda}^2 ( \bd{q} ) \dot{F}_{\Lambda} ( \bd{q} ) F_{\Lambda} ( \bd{q} )  ].
  \label{eq:usharp}
 \end{align}
 \end{subequations}
Choosing a sharp momentum cutoff for simplicity, the
regulated propagators and the corresponding single-scale propagators are 
 \begin{align}
 G_{\Lambda} ( \bd{k} ) & = \frac{ \Theta ( | \bd{k} | - \Lambda ) G_0 ( \bd{k} ) }{
 1 + \Theta ( | \bd{k} | - \Lambda ) G_0 ( \bd{k} ) \Sigma_{\Lambda} ( \bd{k} ) },
 \\
 \dot{G}_{\Lambda} ( \bd{k} ) & = - \frac{ \delta ( | \bd{k} | - \Lambda ) G_0 ( \bd{k} ) }{
 \left[ 1 + \Theta ( | \bd{k} | - \Lambda ) G_0 ( \bd{k} ) \Sigma_{\Lambda} ( \bd{k} ) \right]^2},
 \\
 F_{\Lambda} ( \bd{k} ) & = \frac{ \Theta ( | \bd{k} | - \Lambda ) F_0 ( {\bd{k}} ) }{
 1 + \Theta ( | \bd{k} | - \Lambda ) F_0 ( {\bd{k}} ) \Pi_{\Lambda} ( \bd{k} ) },
 \\
 \dot{F}_{\Lambda} ( \bd{k} ) & = - \frac{ \delta ( | \bd{k} | - \Lambda ) 
F_0 ( {\bd{k}} ) }{
 \left[ 1 + \Theta ( | \bd{k} | - \Lambda ) F_0 ( {\bd{k}} ) \Pi_{\Lambda} ( \bd{k} )  \right]^2}.
 \end{align} 
Substituting these expressions into the flow equations (\ref{eq:flowtrunc}) we obtain various expressions where the $\delta ( | \bd{k} | - \Lambda )$ is multiplied by a function of the step function $\Theta ( | \bd{k} | - \Lambda )$, which are defined via the Morris Lemma \cite{Morris94,Kopietz10}
 \begin{equation} \label{eq:Morris_Lemma}
 \delta ( x ) f ( \Theta ( x ) ) = \delta ( x ) \int_0^1 dt f ( t ).
 \end{equation}
With this prescription the our flow equations~(\ref{eq:flowtrunc}) can be reduced to the following from,
 \begin{subequations}
 \label{eq:flowtrunc2}
 \begin{align}
 \partial_{\Lambda} \Gamma^{(0)}_{\Lambda} & = - \frac{\cal{V}}{2} \int_{\bd{q}}
  \delta ( | \bd{q} | - \Lambda ) \ln \left[
  \frac{ G_0^{-1} ( \bd{q} )+ \Sigma_{\Lambda} ( \bd{q}) }{ G_0^{-1} ( \bd{q} )}
   \right], 
   \\
 \partial_{\Lambda} \sigma_{\Lambda}  & = - \frac{g_{\Lambda}}{2} \int_{\bd{q}} \frac{ \delta ( | \bd {q}  | - \Lambda ) }{ G_0^{-1} ( \bd{q} ) + \Sigma_{\Lambda} ( \bd{q} ) },
 	\\
 \partial_{\Lambda} \Sigma_{\Lambda} (0) & =  - \frac{u_{\Lambda}}{2} \int_{\bd{q}} \frac{ \delta ( | \bd {q}  | - \Lambda ) }{ G_0^{-1} ( \bd{q} ) + \Sigma_{\Lambda} ( \bd{q} ) }
 \nonumber
 \\
 & \hspace{-12mm} +  g_{\Lambda}^2 \int_{\bd{q}} \frac{ \delta ( | \bd{q} | - \Lambda ) }{
  [ G_0^{-1} ( \bd{q} ) + \Sigma_{\Lambda} ( \bd{q} ) ][ F_0^{-1} ( {\bd{q}} ) + \Pi_{\Lambda} ( \bd{q} ) ] },
 \\
 &
 \nonumber
 \\
 \partial_{\Lambda} \Pi_{\Lambda} (0) & =  \frac{g_{\Lambda}^2}{2} 
 \int_{\bd{q}} \frac{ \delta ( | \bd{q} | - \Lambda )}{ [  G_0^{-1} ( \bd{q} ) + \Sigma_{\Lambda} ( \bd{q} ) ]^2 }  ,
 \\
 & \nonumber
 \\
 \partial_{\Lambda} g_{\Lambda} & = \frac{ u_{\Lambda} g_{\Lambda}}{2}  
 \int_{\bd{q}} \frac{ \delta ( | \bd{q} | - \Lambda )}{ [  G_0^{-1} ( \bd{q} ) + \Sigma_{\Lambda} ( \bd{q} ) ]^2 } 
 \nonumber
 \\ &
 \hspace{-12mm} -  g_{\Lambda}^3 \int_{\bd{q}} \frac{ \delta ( | \bd{q} | - \Lambda ) }{
  [ G_0^{-1} ( \bd{q} ) + \Sigma_{\Lambda} ( \bd{q} ) ]^2[ F_0^{-1} (  {\bd{q}} ) + \Pi_{\Lambda} ( \bd{q} ) ] },
 \nonumber
 \\
 &
 \\
 \partial_{\Lambda} u_{\Lambda} & = \frac{ 3 u_{\Lambda}^2}{2}   
 \int_{\bd{q}} \frac{ \delta ( | \bd{q} | - \Lambda )}{ [  G_0^{-1} ( \bd{q} ) + \Sigma_{\Lambda} ( \bd{q} ) ]^2 }
 \nonumber
 \\ & \hspace{-10mm}
 -  6 u_{\Lambda} g_{\Lambda}^2  
  \int_{\bd{q}} \frac{ \delta ( | \bd{q} | - \Lambda ) }{
  [ G_0^{-1} ( \bd{q} ) + \Sigma_{\Lambda} ( \bd{q} ) ]^2[ F_0^{-1} ({\bd{q}} ) + \Pi_{\Lambda} ( \bd{q} ) ] }
 \nonumber
 \\
& \hspace{-10mm} + 6  g_{\Lambda}^4   \int_{\bd{q}} \frac{ \delta ( | \bd{q} | - \Lambda ) }{
  [ G_0^{-1} ( \bd{q} ) + \Sigma_{\Lambda} ( \bd{q} ) ]^2 [ F_0^{-1} ({\bd{q}} ) + \Pi_{\Lambda} ( \bd{q} ) ]^2 }.
 \end{align}
 \end{subequations}
Next, we substitute Eqs.~(\ref{eq:free_phi_propagator}) and (\ref{eq:free_E_propagator})
for $G_0^{-1 }( \bd{q} )$ and $F_0^{-1} ( \bd{q} )$. Because we neglect the momentum dependence
of the self-energies \eqref{eq:SPinull}, the integrations in Eqs.~(\ref{eq:flowtrunc2}) 
can be carried out explicitly using $D$-dimensional spherical coordinates, 
 \begin{subequations}
 \label{eq:flowtrunc3}
 \begin{align}
 \partial_{\Lambda} \Gamma^{(0)}_{\Lambda} & = - \frac{{\cal{V}}}{2}  K_D \Lambda^{D-1}
 \ln \left[ \frac{  r_{\Lambda}  + c_0 \Lambda^2}{ c_0 \Lambda^2} \right],
 \label{eq:flowGammarg}
 \\
 \partial_\Lambda\sigma_{\Lambda} &=-  \frac{  K_D\Lambda^{D-1} }{  r_{\Lambda} + c_0 \Lambda^2  } \frac{g_\Lambda}{2}  , 
 \label{eq:flowsigmarg}
 \\
 \partial_\Lambda \Sigma_{\Lambda} (0) & = \frac{  K_D\Lambda^{D-1} }{  r_{\Lambda} + c_0 \Lambda^2  }\left[- \frac{u_\Lambda}{2} +\frac{g_\Lambda^2}{\rho_{\Lambda}}\right],
 \\
 \partial_\Lambda \Pi_{\Lambda} (0) & =   
  \frac{ K_D \Lambda^{D-1}   }{  ( r_{\Lambda} + c_0 \Lambda^2 )^2}  \frac{ g_{\Lambda}^2}{2},
  \label{eq:Pinullflow}
 \\
 \partial_\Lambda g_{\Lambda} & =  \frac{  K_D\Lambda^{D-1}}{ ( r_{\Lambda} + c_0 \Lambda^2 )^2}\left[\frac{u_\Lambda g_\Lambda}{2}-\frac{g_\Lambda^3}{\rho_{\Lambda}}\right],
 \\
 \partial_\Lambda u_{\Lambda} & = 
  \frac{K_D  \Lambda^{D-1}}{ ( r_{\Lambda} + c_0 \Lambda^2 )^2} \left[\frac{3}{2}u_\Lambda^2-6 \frac{  u_{\Lambda} g_\Lambda^2}{\rho_{\Lambda}}+6\frac{g_\Lambda^4}{\rho_{\Lambda}^2}\right].
 \end{align}  
 \end{subequations}
Here, we have defined
 \begin{subequations}
 \begin{align}
 r_{\Lambda} & = \Sigma_{\Lambda} ( 0 ),
 \\
 \rho_{\Lambda}  & = \rho_0 + \Pi_{\Lambda} (0 ),
 \end{align}   
 \end{subequations}
 and the numerical constant $K_D = \Omega_D / ( 2 \pi )^D$ is the surface area $\Omega_D$ of the $D$-dimensional unit sphere divided by $(2  \pi )^D$.
 For finite external strain $E_0 = {\cal{V}} e_0$, these flow equations should be integrated with the following initial conditions at some large initial scale $ \Lambda = \Lambda_0$,
  \begin{subequations}
   \begin{align}
    \Gamma^{(0)}_{\Lambda_0} & = \frac{ \cal{V}}{2} K_0 e_0^2 ,
     \label{eq:Gammanullinit}
    \\
    \sigma_{\Lambda_0} & = K_0 e_0 ,
    \\
    \Sigma_{\Lambda_0} (0) & =  r_{\Lambda_0} = r_0 + g_0 e_0,
    \label{eq:selfinit}
    \\
    \Pi_{\Lambda_0} (0)& = 0,
    \\
    g_{\Lambda_0} & = g_0,
    \\
    u_{\Lambda_0} & = u_0.
 \end{align}
 \end{subequations}    
Recall that according to Eq.~(\ref{eq:rhodef})
the bulk modulus $K_0 = \rho_0 - 4 \mu /3$ is smaller than the stiffness $\rho_0$.

The simple form of the flow equation (\ref{eq:flowGammarg}) for the free energy is the reason why in Eq.~(\ref{eq:free_phi_propagator}) we have not included the mass term $r_0$ into the definition of $G_0^{-1} ( \bd{k} ) = c_0 k^2$. With the alternative choice
$G_0^{-1} ( \bd{k} ) =  c_0 k^2 + r_0$  the logarithm in the flow equation 
(\ref{eq:flowGammarg}) would be replaced by
$\ln [ ( r_{\Lambda} + c_0 \Lambda^2 )/(r_0 + c_0 \Lambda )]$. 
In Gaussian approximation, where $r_{\Lambda} \approx r_0$, the flow of the free energy would therefore vanish, so that the Gaussian correction to the free energy should be 
taken into account via a non-trivial initial condition. The advantage of
our choice $G_0^{-1} ( \bd{k} ) = c_0 k^2$ is that the Gaussian correction to the free energy is naturally generated by the flow equation (\ref{eq:flowGammarg}).

\section{RG flow for small $\varepsilon = 4 -D$}
\label{sec:fixed4}

\subsection{Field rescaling and anomalous dimensions}
\label{subsec:fieldrescale}

To find the  fixed points of the RG, we introduce the following rescaled  couplings
 \begin{subequations}
 \label{eq:couplings}
 \begin{align}
 r_l & = \frac{ r_\Lambda}{c_\Lambda \Lambda^2 },
 \label{eq:rldef}
 \\
 g_l & = \sqrt{ \frac{ K_D}{ c_{\Lambda}^2 \rho_{\Lambda}  \Lambda^{4-D}} } g_{\Lambda},
 \label{eq:gldef}
  \\
 u_l & = \frac{K_D}{ c_\Lambda^2 \Lambda^{ 4-D}}u_{\Lambda},
 \label{eq:uldef}
 \end{align} 
 \end{subequations}
which are functions of the  logarithmic flow parameter
$l = \ln ( \Lambda_0 / \Lambda )$.  
As usual \cite{Kopietz10}, the coupling 
$c_{\Lambda}$ is determined by the momentum dependence of the Ising self-energy and defines the field rescaling factor $Z_l$ of the Ising field as follows, 
\begin{align}
	Z_{l} & = \frac{c_0}{c_{\Lambda}} =   \frac{1}{1 + c_0^{-1} \left. \frac{ \partial \Sigma_{\Lambda} ( \bd{k} ) }{ \partial ( k^2)  } \right|_{k =0 } }.
	\label{eq:Zdef}
\end{align}
With this definition the term $c_0 k^2$ in the Gaussian part of the action \eqref{eq:Sbare} of the Ising field does not change when we express the action in terms of the rescaled field by setting $ \bar{\phi}_{\bd{k}} = \sqrt{Z_{l}} \phi^{\prime}_{\bd{k}}$.  The logarithmic scale-derivative of the field rescaling factor
$Z_{l}$ defines the scale-dependent anomalous dimension $\eta_l$ of the Ising field,
\begin{align}
 \eta_l & = -  \partial_l \ln Z_l 
 = - \frac{\Lambda}{ c_\Lambda } \lim_{ {k} \rightarrow 0} \frac{ \partial}{\partial ( k^2)  } \partial_\Lambda 
\Sigma_{\Lambda} ( \bd{k} ),
 \label{eq:etadef}
\end{align}
were $\partial_l = - \Lambda \partial_{\Lambda}$. The usual critical exponent
$\eta$ can be identified with $\eta = \lim_{ l \rightarrow \infty} \eta_l$.
Of course, within our truncation $c_\Lambda = c_0 $ and $ \eta_l =0$, because we have neglected the momentum dependence of the Ising self-energy $\Sigma_{\Lambda}$. This is consistent to first order in $\varepsilon = 4 -D$, because $\eta$ is of order  $\varepsilon^2$ \cite{Ma76,Kopietz10}.
On the other hand, the strain field has a non-trivial field-renormalization factor $Y_{\Lambda}$ already at order $\varepsilon$. In analogy with Eq.~(\ref{eq:Zdef}), we define $Y_{\Lambda}$ via
\begin{equation}
	Y_{l} = 
	\frac{ \rho_0 }{ \rho_{\Lambda}  } = \frac{1}{ 1 + \rho_0^{-1}\Pi_{\Lambda} ( 0 ) }
	\label{eq:Ydef}.
\end{equation}
The analogy between Eqs.~(\ref{eq:Zdef}) and (\ref{eq:Ydef}) becomes manifest
by noting  that $\rho_{0}$ is proportional to the square of the longitudinal phonon velocity. Thus, by multiplying the strain self-energy $\Pi_{\Lambda}(0)$ by a factor $k^2$, one can obtain the corresponding self-energy $\tilde{\Pi}(\kv)$ of the phonon displacement field,
 \begin{equation}
	\tilde{\Pi}_{\Lambda} ( {\bd{k}} ) = k^2 \Pi_{\Lambda} ( \bd{k} ).
\end{equation}
Hence, if we demand again that the field rescaling should be chosen such that the  $k^2$ term in the expansion of the inverse phonon propagator is not renormalized, we should choose 
\begin{equation}
	Y_l  = \frac{1}{ 1 +  \rho_{0}^{-1} \left.
		\frac{ \partial \tilde{\Pi}_{\Lambda} ( \bd{k}) }{\partial (k^2)} \right|_{ k =0}},
\end{equation}
which is equivalent to Eq.~(\ref{eq:Ydef}) and
has exactly the same structure as the corresponding expression (\ref{eq:Zdef})
for the Ising field. In analogy with Eq.~(\ref{eq:etadef}), the scale-dependent anomalous dimension $y_l$ of the strain field is therefore  given by
 \begin{align}
 y_l & = - \partial_l \ln Y_l
 = - \frac{ \Lambda  }{ \rho_\Lambda } \partial_{\Lambda}
 \Pi_\Lambda (0).
 \label{eq:ydef}
 \end{align}
Using the definitions (\ref{eq:rldef}, \ref{eq:gldef}, \ref{eq:Ydef}) and 
the flow equation (\ref{eq:Pinullflow}) of  $\Pi_{\Lambda} (0)$ we obtain
 \begin{equation}
 y_l  = - \frac{ g_l^2}{2 ( 1 + r_l )^2 }.
 \label{eq:alphares}
 \end{equation}

\subsection{Fixed points}
 
\begin{figure}[t]
	\begin{center}
		\centering
		\includegraphics[width=0.45\textwidth]{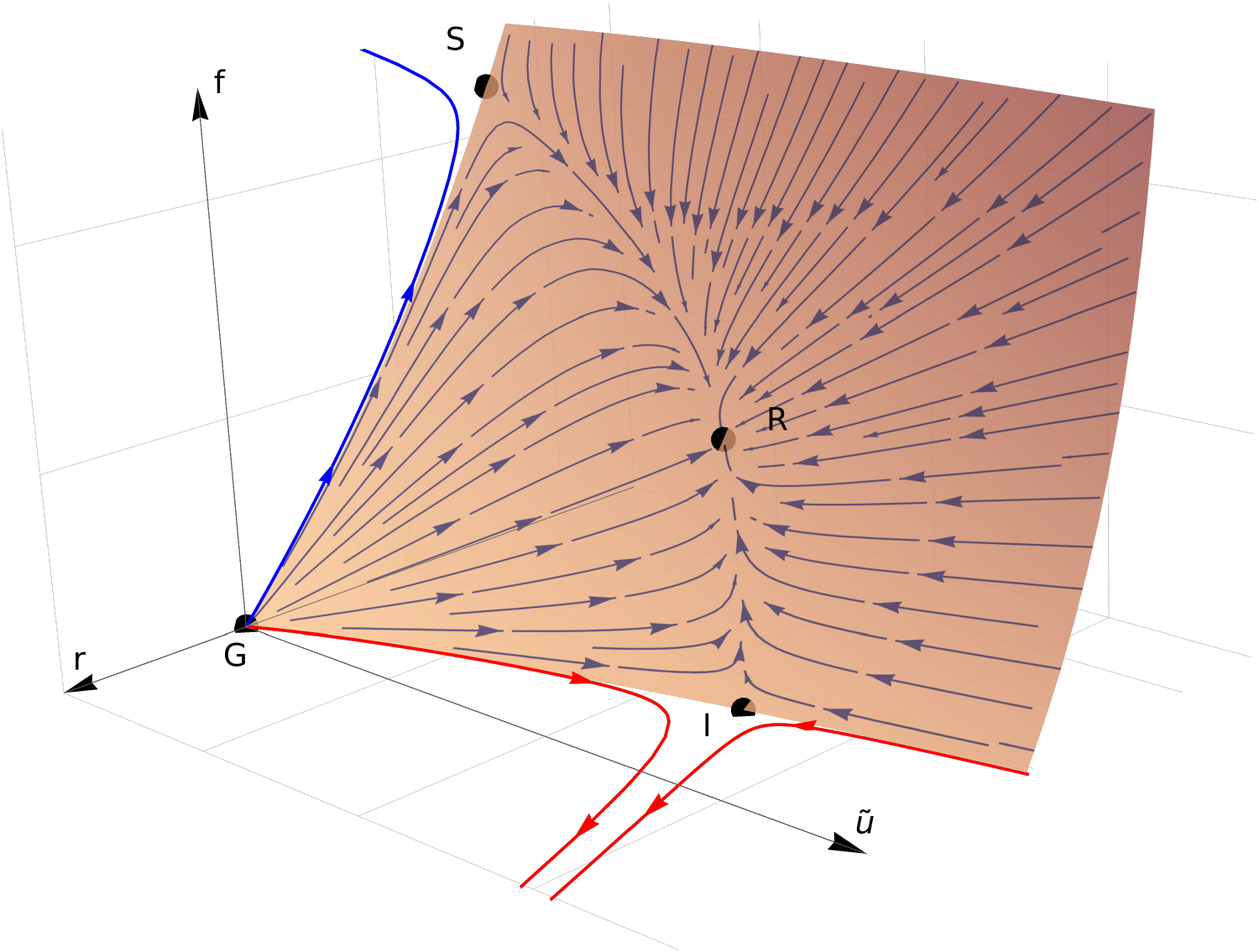}
	\end{center}
	\caption{%
		Graph of the RG flow in the coupling space spanned by $r_l$, $\tilde{u}_{l} = u_l - 3 f_l$, and $f_l = g_l^2$ generated by the flow equations  \eqref{eq:flow_with_ueff}. The shaded surface approximates the critical surface associated with the renormalized Ising fixed point R. The blue and red trajectories illustrate the RG flow originating from initial conditions slightly displaced from the critical surface while passing near the spherical fixed point S and the Ising fixed point I. 
	}
	\label{fig:RGflow}
\end{figure}

Our truncated flow equations (\ref{eq:flowtrunc3}) imply that the rescaled couplings defined 
in Eq.~(\ref{eq:couplings}) satisfy the following flow equations
\begin{subequations}
 \label{eq:flowruv}
 \begin{align}
 \partial_l r_l & = 2 r_l +  \frac{u_l-2g_l^2}{2 ( 1 + r_l )},
 \\
 \partial_l g_l & = \frac{ \varepsilon - y_l}{2} g_{l} - 
  \frac{\left( u_l  -  2 g_l^2\right) g_l}{2( 1 + r_l )^2 }  
  \nonumber
  \\
  & = \frac{\varepsilon }{2} g_l - 
  \frac{\left( 2 u_l  -  5 g_l^2\right) g_l}{4( 1 + r_l )^2 }  ,
  \label{eq:vlflow} 
  \\
 \partial_l u_l & = \varepsilon u_l 
  - \frac{3} {2}\frac{\left(u_l^2 - 4 u_l  g_l^2 + 4 
  	g_l^4\right)}{ ( 1 + r_l )^2} ,
 \label{eq:ulflow}
 \end{align} 
 \end{subequations}
where $\varepsilon = 4- D$ and 
in the last equation we have substituted our result (\ref{eq:alphares}) for $y_l$.
Our flow equations (\ref{eq:flowruv}) are equivalent to the flow equations
obtained by Bergman and Halperin \cite{Bergman76} for the special case of an elastically isotropic system.
Defining
 \begin{subequations}
 \begin{align}
 f_l & = g_l^2, \\
 \tilde{u}_l & = u_l - 3 g^2_l = u_l - 3 f_l,
 \end{align} 
\end{subequations}
the flow equations (\ref{eq:flowruv}) can alternatively be written as
\begin{subequations}\label{eq:flow_with_ueff}
	\begin{align}
		\partial_l r_l & = 2 r_l +  \frac{\tilde{u}_l +f_l}{2 ( 1 + r_l )},
		\\
		\partial_l \tilde{u}_l & = \varepsilon \tilde{u}_l 
		- \frac{ 3 \tilde{u}_l^2}{ 2 ( 1 + r_l )^2} ,
		\\
		\partial_l f_l & = \varepsilon f_l- \frac{\left(2\tilde{u}_l + f_l\right) f_l}{2( 1 + r_l )^2 } . 
	 \end{align} 
\end{subequations}
The advantage of this parametrization is that the flow of the coupling
$\tilde{u}_l$ is decoupled from the flow of the other interaction parameter $f_l$.
The RG flow generated by Eqs.~(\ref{eq:flow_with_ueff}) is shown graphically
in Fig.~\ref{fig:RGflow}.
By demanding that the left-hand sides of Eq.~(\ref{eq:flow_with_ueff}) vanish we find that the  couplings
$r_{\ast}$, $\tilde{u}_{\ast}$, and $f_{\ast}$ at possible RG fixed points must satisfy
\begin{subequations} 
 \label{eq:ruffix}
\begin{align}
   2r_{\ast} & = -  \frac{  \tilde{u}_{\ast} +  f_{\ast}}{ 2 ( 1+ r_{\ast} )},
 \label{eq:rfix}
 \\
 \varepsilon \tilde{u}_{\ast} & = \frac{3 \tilde{u}_{\ast}^2 }{2 ( 1 + r_{\ast} )^2 } ,
 \label{eq:ufix}
 \\
 \varepsilon f_{\ast} & =  \frac{  ( 2 \tilde{u}_{\ast} + f_{\ast} ) f_{\ast}}{2 ( 1 + r_{\ast} )^2}.
 \label{eq:ffixe}
 \end{align}
 \end{subequations}
As anticipated, we obtain for $\varepsilon > 0$ four distinct fixed points
labeled in Fig.~\ref{fig:RGflow}
by G (Gaussian), I (Ising), R (renormalized Ising), and S (spherical).
Although the fixed-point equations (\ref{eq:ruffix}) can  be solved analytically for 
arbitrary $\varepsilon$, in Table
\ref{tab:fixedpoints} we give the couplings at the fixed points only to first order in $\varepsilon$, because our truncation is only accurate to this order.
\begin{table}
\caption{ \label{tab:fixedpoints} Values of the rescaled couplings
$r_\ast$, $u_\ast$, $\tilde{u}_{\ast} = u_{\ast} - 3 f_{\ast}$, $f_{\ast}= g_{\ast}^2$, and the anomalous dimension $y_{\ast}$
of the strain field at the fixed points of the system (\ref{eq:flow_with_ueff})
of flow equations to first order in $\varepsilon = 4 -D > 0$.}
\begin{ruledtabular}
\begin{tabular}{c|c|c|c|c|c}
fixed point  & $r_{\ast}$ \; & $u_{\ast}$ \;  &
$\tilde{u}_{\ast} \; $
 & $f_\ast = g_{\ast}^2$ \;  & $y_{\ast}$ \;  \\
\hline \hline
G (Gaussian)  & $0$ \;   & $0$\;   & $0$ \;  & $0$ \; & $0$\; \\
\hline
I (Ising) & $-  \varepsilon /6  $ \;   & $2  \varepsilon /3 $ \;  & $2 \varepsilon /3$ \;   & $0$ \; & $0$ \;\\
\hline 
R (renorm. Ising)  & $-  \varepsilon /3 $ \;  &  $  8 \varepsilon /3 $ \;  & $ 2 \varepsilon /3 $ \;  & $ 2 \varepsilon /3 $ \; & $- \varepsilon /3 $ \;\\
\hline
S (spherical)  & $-  \varepsilon /2 $ \;  &  $6 \varepsilon$ \; 
 & $ 0$ \; & $2 \varepsilon$ \; & $-\varepsilon$ \; \\
 \end{tabular}
 \end{ruledtabular}
 \end{table}
 
An important fact which, to our best knowledge, has not been noticed before is that the fixed points R and S are characterized by
finite anomalous dimensions $y_\ast$ of the strain field.
As a consequence, in the vicinity of R and S the renormalized stiffness
$\rho_{\ast} ( \bd{k} )$ for strain fluctuations 
is singular and
exhibits anomalous scaling with exponent $y_{\ast}$,
 \begin{equation}
 \rho_\ast ( \bd{k} ) \propto \rho_0 /  Y_{\Lambda = k }
\propto k^{- y_{\ast}} = k^{  | y_\ast | }.
 \end{equation}
Because  $\rho_{\ast} ( \bd{k} )$ is proportional to the
square of the longitudinal sound velocity, we conclude that at the critical points associated with R and S the dispersion of longitudinal sound is, for small momenta, proportional to $ k^{ 1 - y_{\ast}/2}$ which vanishes faster than linear because $y_{\ast} < 0$. 
Considering that sound can be identified with the Goldstone mode associated with the breaking of translational invariance in a crystal, we conclude that the fixed points R and S are examples of a scenario where the dispersion of a Goldstone mode is modified by critical fluctuations.

\subsection{Stability, scaling variables, and Fisher renormalization}
\label{sec:Stabiliy_epsilon}

To analyze the stability of the fixed points 
and classify the scaling variables and the associated eigenvalues, we next linearize the RG flow around the fixed points. Introducing the deviations of the couplings from the fixed-point values,
 $\delta r_l = r_l - r_{\ast}$, $\delta \tilde{u}_l = \tilde{u}_l - \tilde{u}_{\ast}$ and
$\delta f_l = f_l - f_{\ast}$, and linearizing the flow equations~(\ref{eq:flow_with_ueff}) around any given  fixed point, we obtain to first order in $\varepsilon$,
\begin{widetext}
 \begin{align}
	&  \partial_l \begin{pmatrix}
		\delta r_{l}\\
		\delta \tilde{u}_{l}\\
		\delta f_l \\
	\end{pmatrix}  =	\begin{pmatrix}
		2-\bar{u}_{\ast}+f_{\ast} &\frac{1-r_{\ast}}{2} & \frac{1-r_{\ast}}{2} \\
		0  & \varepsilon-3\bar{u}_{\ast} & 0 \\
		0  & f_{\ast} & \varepsilon-\bar{u}_{\ast}-f_{\ast}
	\end{pmatrix}
	\begin{pmatrix}
		\delta r_{l}\\
		\delta \tilde{u}_{l}\\
		\delta f_l \\
	\end{pmatrix} .
	\label{eq:linearized_flow}
	\vspace{0cm}
\end{align}
\end{widetext}
By substituting the fixed-point values from Table~\ref{tab:fixedpoints} into the  matrix in Eq.~\eqref{eq:linearized_flow}, we can calculate the scaling variables $s_{l}$  
from the left-eigenvectors of the matrix and identify the scaling exponents
$\lambda_i$ with the corresponding eigenvalues.
The Gaussian fixed point has  only positive eigenvalues,
 \begin{subequations}
 \begin{align}
 \lambda^G_1 & =  2, 
 \\ 
 \lambda^G_2 & =  \varepsilon , \\
 \lambda^G_3 & = \frac{\varepsilon}{2} .
 \end{align}
 \end{subequations}
The associated scaling variables are
  \begin{subequations}
 \begin{align}
  s^G_{1,l} & = ( 4 - 2 \varepsilon ) \delta r_l + \delta \tilde{u}_l ,
   \\
  s^G_{2,l}  & = \delta \tilde{u}_l , 
  \\ 
  s^G_{3,l} & = \delta f_l.
   \end{align}
   \end{subequations}
For the Ising  fixed point, we find the eigenvalues of the linearized flow
\begin{subequations}
 \begin{align}
 \lambda^I_1 & =  2 - \frac{\varepsilon }{3}, 
 \\ 
 \lambda^I_2 & =  - \varepsilon , \\
 \lambda^I_3 & = \frac{\varepsilon}{3} ,
 \end{align}
 \end{subequations}
and the associated scaling variables
  \begin{subequations}
 \begin{align}
  s^I_{1,l} & = ( 4 - 2 \varepsilon ) \delta r_l + \left( 1 - \frac{2 \varepsilon}{ 3 }
   \right) \delta \tilde{u}_l  + \delta f_l,
   \\
  s^I_{2,l}  & = \delta \tilde{u}_l , 
  \\ 
  s^I_{3,l} & = \delta f_l.
   \end{align}
   \end{subequations}
As graphically depicted in Fig.~\ref{fig:RGflow}, the eigenvalues around the renormalized Ising fixed point have only one relevant eigenvalue
\begin{subequations}
 \begin{align}
 \lambda^R_1 & =  2 - \frac{2 \varepsilon }{3}, 
 \\ 
 \lambda^R_2 & =  - \varepsilon , \\
 \lambda^R_3 & = - \frac{\varepsilon}{3} .
 \end{align}
 \end{subequations}
The associated scaling variables are
  \begin{subequations}
 \begin{align}
  s^R_{1,l} & = ( 4 - 2 \varepsilon ) \delta r_l + \left( 1 - \frac{2 \varepsilon}{ 3 }
   \right) \delta \tilde{u}_l ,
   \\
  s^R_{2,l}  & = \delta \tilde{u}_l , 
  \\ 
  s^R_{3,l} & = - \delta \tilde{u}_l + \delta f_l.
   \end{align}
   \end{subequations}
Finally, the spherical fixed point is characterized by the eigenvalues
\begin{subequations}
 \begin{align}
 \lambda^S_1 & =  2 - 2 \varepsilon, 
 \\ 
 \lambda^S_2 & =   \varepsilon , 
 \\
 \lambda^S_3 & =  - \varepsilon ,
 \end{align}
 \end{subequations}
and  the associated scaling variables
  \begin{subequations}
 \begin{align}
  s^S_{1,l} & = ( 4 - 4 \varepsilon ) \delta r_l +  \delta \tilde{u}_l + \delta f_l,
   \\
  s^S_{2,l}  & = \delta \tilde{u}_l , 
  \\ 
  s^S_{3,l} & =  \delta \tilde{u}_l + \delta f_l.
   \end{align}
   \end{subequations}
The eigenvalue $\lambda_1$, associated with the thermal scaling variable
$s_{1 , l }$, determines the correlation length exponent $\nu = 1/ \lambda_1$.
To leading order in $\varepsilon$, we find that our four distinct fixed points are characterized by
\begin{subequations}
	\begin{align}
		\nu_G&=\frac{1}{2}, \\
		\nu_I&=\frac{1}{2-\frac{\varepsilon}{3}}\approx \frac{1}{2}+\frac{\varepsilon}{12}, \\
		\nu_R&=\frac{1}{2-\frac{2}{3}\varepsilon}\approx \frac{1}{2}+\frac{\varepsilon}{6},\\
		\nu_S &=\frac{1}{2-2\varepsilon}\approx \frac{1}{2}+\frac{\varepsilon}{4}.
	\end{align}
\end{subequations}
Using the fact that for $D < 4$ the  specific heat exponent $\alpha$
is related to $\nu$ via the hyperscaling relation
\begin{equation}\label{eq:hyper_scaling_alpha}
	\alpha = 2 - D \nu,
\end{equation}
we obtain the corresponding specific heat exponent to leading order in $\varepsilon$
\begin{subequations}
	\begin{gather}
		\alpha_G =\frac{\varepsilon}{2},\\
		\alpha_I \approx \frac{\varepsilon}{6},\\
		\alpha_R \approx -\frac{\varepsilon}{6},\\
		\alpha_S \approx -\frac{\varepsilon}{2}.
	\end{gather}
\end{subequations}
As first noted by Bergman and Halperin \cite{Bergman76}, these relations are compatible with
Fisher renormalization \cite{Fisher68} up to order 
$\varepsilon$, which implies the following relations between the specific heat exponents,
\begin{subequations}
	\begin{gather}
		\alpha_S=\frac{-\alpha_G}{1-\alpha_G}, \label{eq:Fischer_rescaling_S}\\
		\alpha_R=\frac{-\alpha_I}{1-\alpha_I}. \label{eq:FIscher_rescaling_R}
	\end{gather}
\end{subequations}
The reason why R is called the renormalized Ising fixed point is that
the critical exponents of I and R are related by 
Fisher renormalization \cite{Fisher68}.
Moreover, since $\lambda_1^G=2$ is exact at the Gaussian fixed point, we can use  
Eq.~\eqref{eq:Fischer_rescaling_S} in combination with the hyperscaling 
relation~\eqref{eq:hyper_scaling_alpha} to obtain the exact specific heat exponent 
$\alpha_S$ of the spherical model \cite{Berlin52,Stanley68}
\begin{equation}\label{eq:exact_spherical}
	\alpha_S=-\frac{2-\frac{D}{2}}{1-\left(2-\frac{D}{2}\right)}=-\frac{4-D}{D-2}.
\end{equation}
This is the reason why, for our model, the fixed point $S$ is called the 
spherical fixed point \cite{Rudnick74}.

\section{RG flow in three dimensions}
\label{sec:FRG3}

A priori, it is not clear whether the fixed points obtained for small $\varepsilon $ persist in three dimensions. For example, using non-perturbative truncations of the FRG flow equations of $D$-dimensional $O(N)$ models, Yabunaka and Delamotte \cite{Yabunaka17} demonstrated that in three dimensions these models have several non-perturbative fixed points which are missed by the $\varepsilon$-expansion. 
Moreover, as a function of the dimensionality $D$ and the number $N$ of field components, new fixed points can appear or fixed points can merge and disappear.
The truncation of the FRG flow equations adopted 
in Sec.~\ref{sec:fixed4} is quantitatively accurate only  
to leading order in $\varepsilon = 4-D$. In particular, we have 
retained only the vertices $\Gamma^{\phi \phi \epsilon}_{\Lambda}$ and
$\Gamma^{\phi \phi \phi \phi}_{\Lambda}$, which appear already in the bare action, and have neglected the momentum dependence of all vertices, including the
two-point vertices. 
In three dimensions this truncation
is not well justified, because at the fixed points the vertices which we have omitted
in the truncated vertex expansion (\ref{eq:vertexp}) are expected to have the same order of magnitude as the vertices which we have retained. To go beyond this approximation within the FRG framework, we can adopt a more general ansatz for the average effective action
$\Gamma_{\Lambda} [ \bar{\phi} , \bar{\epsilon} ]$ that is consistent with the symmetries of the system  \cite{Wetterich93,Berges02,Pawlowski07,Kopietz10,Dupuis21}.
For simplicity, in this section we set the uniform strain $E_0$ equal to zero.
Our improved truncation retains all vertices with three and four external legs allowed by symmetry. 
The corresponding ansatz for the generating functional of the truncated average effective action is
\begin{widetext}
 \begin{eqnarray}
 \Gamma_{\Lambda} [ \bar{\phi} , \bar{\epsilon} ] & = & \Gamma_{\Lambda}^{(0)} + \frac{1}{2} \int_{\bd{k}}
 \left[  G_0^{-1} ( \bd{k} ) + \Sigma_{\Lambda} ( \bd{k} ) \right]
 \bar{\phi}_{ - \bd{k}} \bar{\phi}_{\bd{k}} 
 + \frac{1}{2} \int_{\bd{k}} \left[ F_0^{-1} ( {\bd{k}} ) + \Pi_{\Lambda} ( \bd{k} ) \right]
 \bar{\epsilon}_{ - \bd{k}} \bar{\epsilon}_{\bd{k}} 
 \nonumber
 \\
 & + &   \int_{\bd{k}_1} \int_{\bd{k}_2} \int_{\bd{k}_3}
 {\cal{V}} \delta_{  \bd{k}_1 + \bd{k}_2 + \bd{k}_3 ,0}
 \left[ \frac{1}{2!}
 \Gamma_{\Lambda}^{\phi \phi \epsilon} ( \bd{k}_1 , \bd{k}_2 , \bd{k}_3 )
  \bar{\phi}_{\bd{k}_1 }  \bar{\phi}_{\bd{k}_2} \bar{\epsilon}_{\bd{k}_3}
 +
 \frac{1}{3!} \Gamma_{\Lambda}^{\epsilon \epsilon \epsilon} ( \bd{k}_1 , \bd{k}_2 , \bd{k}_3 )
  \bar{\epsilon}_{\bd{k}_1 }  \bar{\epsilon}_{\bd{k}_2} \bar{\epsilon}_{\bd{k}_3}
 \right]
 \nonumber
 \\
  & + & \int_{\bd{k}_1}  \int_{\bd{k}_2}   \int_{\bd{k}_3}    \int_{\bd{k}_4}
 {\cal{V}} \delta_{ \bd{k}_1 + \bd{k}_2 + \bd{k}_3 + \bd{k}_4 ,0} 
 \biggl[
  \frac{1}{4!}
\Gamma^{\phi \phi \phi \phi}_{\Lambda}
 ( \bd{k}_1 , \bd{k}_2 , \bd{k}_3 , \bd{k}_4 )
 \bar{\phi}_{\bd{k}_1 } \bar{\phi}_{\bd{k}_2} \bar{\phi}_{\bd{k}_3} \bar{\phi}_{\bd{k}_4} 
 \nonumber
 \\
 & &
 +  \frac{1}{4!}
\Gamma^{\epsilon \epsilon \epsilon \epsilon}_{\Lambda}
 ( \bd{k}_1 , \bd{k}_2 , \bd{k}_3 , \bd{k}_4 )
 \bar{\epsilon}_{\bd{k}_1 } \bar{\epsilon}_{\bd{k}_2} \bar{\epsilon}_{\bd{k}_3} 
 \bar{\epsilon}_{\bd{k}_4}
 +  \frac{1}{(2!)^2}
\Gamma^{\phi \phi \epsilon \epsilon}_{\Lambda}
 ( \bd{k}_1 , \bd{k}_2 , \bd{k}_3 , \bd{k}_4 )
 \bar{\phi}_{\bd{k}_1 } \bar{\phi}_{\bd{k}_2} \bar{\epsilon}_{\bd{k}_3} 
 \bar{\epsilon}_{\bd{k}_4}
 \biggr].
 \label{eq:vertexp2}
 \end{eqnarray}
 \end{widetext}
In the spirit of the local-potential approximation \cite{Berges02}, we neglect the momentum dependence of all vertices. Apart from the two couplings which we have already taken into account close to four dimensions,
 \begin{subequations}
 \label{eq:gu} 
 \begin{align}
 g_{\Lambda} & = \Gamma_{\Lambda}^{\phi \phi \epsilon} ( 0,0,0),
 \\
 u_{\Lambda} & = \Gamma_{\Lambda}^{\phi \phi \phi \phi} (0,0,0,0),
 \end{align}
 \end{subequations}
we now also include the RG flow of  
 \begin{subequations}
 \label{eq:uvw} 
 \begin{align}
 h_{\Lambda} & =   \Gamma_{\Lambda}^{\epsilon \epsilon \epsilon} ( 0,0,0),
 \\
 v_{\Lambda} & =  \Gamma_{\Lambda}^{\epsilon \epsilon \epsilon \epsilon} ( 0,0,0,0),
 \\
 w_{\Lambda} & = \Gamma_{\Lambda}^{\phi \phi \epsilon \epsilon} (0,0,0,0),
 \end{align}
  \end{subequations}
where we have used the same notation as in Eqs.~(\ref{eq:Sanharm}) and (\ref{eq:Shigh}). 
Writing down coupled truncated FRG flow equations
for the above couplings is now straightforward. Nevertheless, we present the details in Appendix~B because calculations are rather lengthy. The main result of this calculation is that all four fixed points
G, I, R, and S survive in three dimensions. However, in the extended coupling space spanned by $g_l, u_l, h_l, v_l,$ and $w_l$, the new 
flow equations feature additional fixed points which we believe to be artifacts of our truncation and therefore discard.

\section{Elastic equation of state, bulk instability, and Hooke's law}
\label{sec:free_energy}

In this section, we use our FRG formalism to
calculate the elastic equation of state, i.e., 
the stress $\sigma$ as a function of the homogeneous  
strain $e_0$. For simplicity, we use the truncation of the FRG flow equations developed in Sec.~\ref{sec:FRG4}, which is
quantitatively accurate to leading order in $\varepsilon$.
Given the fact that in our bare action $S [ \phi, E ]$ defined in Eq.~(\ref{eq:Sbare}) the homogeneous part $E_0 = {\cal{V}} e_0$ of the strain field can be identified with the volume strain, the conjugate homogeneous stress $\sigma(e_0)$ can be identified with the hydrostatic pressure. 
Since we work at fixed strain $e_0$, the conjugate stress $\sigma_{\Lambda}$
changes with the RG scale $\Lambda$. The corresponding flow equation (\ref{eq:flowsigmarg})
derived in Sec.~\ref{sec:FRG4} relates the change of $\sigma_{\Lambda}$ to
the scale-dependent couplings $r_{\Lambda}$ and $g_{\Lambda}$.
Introducing the rescaled stress
 \begin{align}
 \sigma_{l} & =	\frac{\sigma_{\Lambda}}{ \left(\rho_{\Lambda}K_D\Lambda^{D}\right)^{1/2}},
 \label{eq:sigmaLL}
 \end{align} 
the flow equation (\ref{eq:flowsigmarg}) can be expressed in terms of our rescaled couplings as follows,
 \begin{align}
 \partial_l \sigma_{l} &=\frac{D-y_l}{2} \sigma_{l}+ \frac{g_l}{2( 1 + r_l)}.
 \label{eq:sigmalflow}
 \end{align}
Integrating this over the logarithmic flow parameter $l$, substituting the result 
into (\ref{eq:sigmaLL}), and finally taking the limit $\Lambda \rightarrow 0$ we can express the  physical stress $\sigma = \lim_{\Lambda \rightarrow 0} \sigma_{\Lambda}$ 
in terms of  the flow of the rescaled couplings $r_l$ and $g_l$ on the entire RG trajectory,
 \begin{align}
 \sigma & = K_0 e_0  
 \nonumber
 \\
 & + \sqrt{K_D \Lambda_0^D \rho_0 } 
 \int_0^\infty dl  e^{ - \frac{1}{2} \int_0^l d t ( D - y_t ) } \frac{g_l}{2 ( 1 + r_l ) }.
 \end{align}
Alternatively, we can obtain the stress $\sigma$ by first calculating the free energy 
$\Gamma^{(0)} = \lim_{\Lambda \rightarrow 0} \Gamma^{(0)}_{\Lambda}$  at fixed strain $e_0$ and then differentiating with respect to $e_0$,
 \begin{equation}
 \sigma = \frac{1}{ \cal{V}} \frac{ \partial \Gamma^{(0)} }{\partial e_0 }.
 \label{eq:stressGamma0}
 \end{equation}
Integrating the flow equation (\ref{eq:flowGammarg}) over the cutoff parameter $\Lambda$ we obtain
for the free energy per volume at constant strain,
 \begin{align} 
  \frac{\Gamma^{(0)}}{\cal{V}} & =   
  \frac{K_0}{2} e_0^2 + \frac{K_D}{2}  \Lambda_0^D \int_0^\infty dl e^{ - D l }
  \ln \left( \frac{ 1 + r_l}{Y_l}   \right),
  \label{eq:Gammaadv}
  \end{align}
where we already inserted the rescaling factor $Y_l$ of the strain field, as defined in Eq.~(\ref{eq:Ydef}).

We are interested in the scaling properties of the free energy 
in the vicinity of the four fixed points G, I, R, and S 
discussed in Sec.~\ref{sec:FRG4}.
Assuming that the RG trajectory lies close to a fixed point, we may approximate the  flow of $r_l$ close to the fixed point $r_{\ast}$ by the linearized flow 
$r_l = r_{\ast} + \delta r_l$, where the deviation $\delta r_l$ from the fixed point
obeys the linearized flow equation (\ref{eq:linearized_flow}). Similarly, we approximate the scale-dependent anomalous dimension $y_l$ defined in  Eq.~(\ref{eq:ydef}) 
  via the logarithmic scale derivative of $Y_l$ by its fixed-point value
  $y_{\ast}$ so that
  \begin{equation}\label{eq:Y_approx}
	\ln (1/Y_{l}) \approx \int_{0}^{l}dl'y_{\ast}= y_{\ast}l.
\end{equation}
In this approximation
\begin{align} 
  & \frac{\Gamma^{(0)}}{\cal{V}}  \approx   
  \frac{K_0}{2} e_0^2 
 \nonumber
  \\
  &  
  + \frac{K_D}{2}  \Lambda_0^D \int_{l_c}^{l_{\ast}} dl e^{ - D l }
  \left[ \ln \left(  1 +  r_{\ast} + \delta r_l \right) + y_{\ast} l \right].
  \label{eq:Gammaadv}
  \end{align}
Here, the lower limit is determined by the Ginzburg scale
$ \Lambda_c = \Lambda_0 e^{ - l_c }$, while the upper limit is given by the
inverse correlation length $\Lambda_{\ast} = \Lambda_0 e^{ - l_{\ast} }$, and we have assumed that $\Lambda_{\ast} \ll \Lambda_c$ so that $l_{\ast} - l_c \gg 1$.
From the stability analysis of the linearized flow in the vicinity of the fixed points 
presented in Sec.~\ref{sec:Stabiliy_epsilon} we know, at least for small $\varepsilon$, that
the scaling in the vicinity of all fixed points is dominated by the relevant thermal 
scaling field $ s_{1, l} \propto \delta r_l$. Because for small $\varepsilon$  the  associated eigenvalue $\lambda_1$ is  large compared with the eigenvalues of all other scaling variables, we can approximate
\begin{equation}
	\delta r_l = t e^{\lambda_1 l},
	\label{eq:rtldef}
\end{equation}  
where the dimensionless parameter $t$ is defined by
 \begin{equation}
 t = a ( T - T_c )  + b  g_0 e_0  = t_0 + \tilde{g}_0 e_0 .
 \end{equation}
Here $a$ and $b$ are dimensionful constants, $t_0 = a ( T - T_c)$,
$\tilde{g}_0 = b g_0$, and $T$ denotes the temperature, while $T_c$ is the critical temperature at $e_0=0$. 
Note that the scale  $l_{\ast}$ in Eq.~(\ref{eq:Gammaadv}) can be  defined by
 \begin{equation}
 1 = t e^{ \lambda_1 l_{\ast} },
 \label{eq:txi}
 \end{equation}
where $e^{l_{\ast}}$ can be identified with the correlation length $\xi$ in units of 
an underlying lattice spacing. Hence,  Eq.~(\ref{eq:txi})
is equivalent with $\xi \propto t^{-1/ \lambda_1 } \equiv t^{ - \nu }$,
where $\nu = 1 / \lambda_1$ is the correlation length exponent.
Substituting Eq.~(\ref{eq:rtldef}) into the free energy Eq.~(\ref{eq:Gammaadv}) and introducing the new integration variable $x =  t e^{ \lambda_1 l}$ we obtain
 \begin{align} 
   \frac{\Gamma^{(0)}}{\cal{V}}  \approx   
  \frac{K_0}{2} e_0^2   + & \frac{K_D}{2}  \Lambda_0^D \nu  t^{D \nu} 
   \int_{ c t  }^{1} dx x^{ - D \nu   -1 } 
 \nonumber 
  \\
  &  
  \times
  \left[ \ln \left(  1 +  r_{\ast} + x \right) + y_{\ast} \ln ( x/t) \right],
  \label{eq:Gamma0scale}
  \end{align}
where the numerical constant $c = e^{ \lambda_1 l_c }$  will be 
replaced by unity from now on.  
Repeatedly integrating by parts and using 
the hyperscaling relation (\ref{eq:hyper_scaling_alpha}) 
to write $D / \lambda_1 = D \nu = 2 - \alpha$
we can  cast Eq.~(\ref{eq:Gamma0scale}) 
into the following form
 \begin{align} 
   \frac{\Gamma^{(0)}}{\cal{V}} & \approx   
  \frac{K_0}{2} e_0^2   +  A_0 + A_1 t + \frac{A_2}{2}  t^2    
 \nonumber
  \\  
  & -   A t^{2 - \alpha}  +   y_{\ast}   B  t^{2 - \alpha}  \ln t   + {\cal{O}} ( t^3 ) ,
  \label{eq:Gamma1scale}
  \end{align}
where 
$B > 0$ and, for small $\varepsilon$, the constants  $A_2$ and $A$ are 
positive if $\alpha >0$ and  negative if $\alpha < 0$. 
Using Eq.~(\ref{eq:stressGamma0}) to obtain the stress from the strain-derivative of the free energy we obtain
 \begin{align}
 \sigma  - \sigma_0 & \approx K_0 e_0 +   \tilde{g}_0 \Bigl[ A_2 t -  [ ( 2 - \alpha ) A - y_{\ast} B ] t^{1 - \alpha} 
  \nonumber
  \\
  & + ( 2 - \alpha ) y_{\ast} B t^{1- \alpha} \ln t  + {\cal{O}} ( t^2 )
 \Bigr],
 \label{eq:eleq}
 \end{align}
where $\sigma_0 = \tilde{g}_0 A_1$.
Taking one more derivative with respect to the strain $e_0$, we obtain the renormalized bulk modulus
\begin{align}
	K (t) & = K_0 +  A_2\tilde{g}_0^2 \nonumber\\
	&-\tilde{g}_0^2\left[ ( 2 - \alpha )( 1 - \alpha ) A -  (3 - 2 \alpha) y_{\ast} B \right] 	t^{-\alpha} \nonumber  \\
	&+ \tilde{g}_0^2\left[( 2 - \alpha ) ( 1 - \alpha) y_{\ast} B\right]  t^{- \alpha}  \ln t + {\cal{O}} ( t ).
\label{eq:Kren}
\end{align} 

First, consider the Ising fixed point I where $y_{\ast} =0$ and $\alpha > 0$. For small $t$, the renormalized bulk modulus is then dominated by the singular contribution proportional to $t^{- \alpha}$. Since $A > 0$ for positive $\alpha$, 
the renormalized bulk modulus in Eq.~(\ref{eq:Kren}) vanishes at a finite value of the reduced temperature $t$. Physically, this means that Ising criticality is preempted by a bulk instability before the critical point can be reached~\cite{Bergman76,Zacharias12,Zacharias15,Zacharias15b}.
In Ref.~[\onlinecite{Zacharias12}] Zacharias et al.  have related such an isostructural instability (where $K \rightarrow 0$) to the breakdown of Hooke's law of elasticity, leading to a non-linear stress-strain relation characterized by a mean-field exponent $\delta=3$. 
However, in Ref.~[\onlinecite{Zacharias12}] the authors have considered a 
system in a finite magnetic field $H$ which gives rise to a 
hybridization $g_{H}\phi(\mathbf{r})E(\mathbf{r})$
between Ising- and elastic fluctuations. In this case
the magnetic scaling field close to the Ising fixed point is proportional to $  H+g_{H}e_0$ and can therefore be tuned by the external strain $e_0$. In contrast, here we consider only the case of zero magnetic field where $g_H =0$ and the exponent
$\delta$ does not play any role for the breakdown of Hooke's law.
%

Next, let us consider the elastic equation of state (\ref{eq:eleq})
at the 
renormalized Ising fixed point R  and at the spherical fixed point S.
In this case both exponents $y_{\ast}$ and $\alpha$ are negative.
For simplicity, we focus on the elastic equation of state (\ref{eq:eleq})
at the critical temperature so that we set $t_0=0$ and, hence, $t = \tilde{g}_0 e_0$. Retaining in Eq.~(\ref{eq:eleq}) only the dominant linear term and the leading non-analytic correction, we obtain
 \begin{align}
  \sigma - \sigma_0 & \approx ( K_0 + \tilde{g}_0^2 A_2 ) e_0 
  \nonumber
   \\
   & + \tilde{g}_0^{2 - \alpha}  ( 2 - \alpha) y_{\ast} B e_0^{1- \alpha} \ln ( \tilde{g}_0 e_0 ).
   \label{eq:stressstrain}
  \end{align}
Because $\alpha < 0$ at R and S, the linear term dominates the elastic equation of state for sufficiently small $e_0$, provided the renormalized bulk modulus
 \begin{equation}
 K = K_0 + \tilde{g}_0^2 A_2
 \label{eq:KK}
 \end{equation}
is positive, thereby ensuring thermodynamic stability. Although $A_2$ is negative for small $\varepsilon$, where our calculation is controlled, the renormalized bulk modulus remains positive as long as the bare coupling $g_0$ is sufficiently small. This regime, in which the stress-strain relation is predominantly linear for small $e_0$, is commonly referred to as Hooke's law. However, the leading correction to Hooke's law
is non-analytic  and proportional to $ e_0^{1 - \alpha} | \ln ( e_0 ) |$ at $t_0=0$, 
where the logarithmic correction is due to the anomalous dimension of the strain field
at the fixed points R and S. In three dimensions, we may combine the known 
value $\alpha_I \approx 0.11$ of the specific heat exponent of the Ising universality class \cite{Pelissetto02} with Fisher renormalization (\ref{eq:FIscher_rescaling_R}) to estimate $\alpha_R \approx -0.12$ so that  the leading non-analytic correction to Hooke's law at the
renormalized Ising fixed point is of order $e_0^{1.12} | \ln ( e_0 ) |$.
On the other hand, using the explicit expression (\ref{eq:exact_spherical}) for the specific heat exponent $\alpha_S$ of the spherical model we obtain in three dimensions  $\alpha_S = -1$ so that the leading non-analytic correction to Hooke's law is proportional to $e_0^2 | \ln e_0 |$.
Because for small $\varepsilon$ the coefficient
$A_2 $ in Eq.~(\ref{eq:KK}) is negative,
the renormalized compressibility $K$ in Eq.~(\ref{eq:KK}) vanishes at $\tilde{g}_0^2 = - K_0 /  A_2$, signaling again a bulk instability. At this point, Hooke's law breaks down and the non-analytic term determines the stress-strain relation in the second line of Eq.~(\ref{eq:stressstrain}).

\section{Summary and conclusions}
\label{sec:summary_and_discussion}

In this work, we have re-examined the effect of elastic fluctuations on Ising criticality using a modern functional renormalization group approach, which is implemented such that the strain, and therefore the volume, is held fixed during the RG flow.
For small  $\varepsilon = 4 -D$ we have recovered the four distinct fixed points G, I, R, and S of the constrained Ising model obtained previously
by Bergman and Halperin \cite{Bergman76} by employing a perturbative momentum-shell RG procedure controlled to first order in $\varepsilon$.
Using a more sophisticated truncation of the FRG flow equations including all three-point and four-point vertices allowed by symmetry, we have presented evidence that these fixed points survive down to three dimensions.

In general, fixed points of RG flows can only be obtained if the
relevant fields are rescaled. The proper choice of the field rescaling factors is rather subtle and depends on the specific field considered, the symmetry of a model, and the type of fixed points. In Sec.~\ref{subsec:fieldrescale} we have carefully derived the proper
rescaling factor $Y_l$ of the strain field in our model Eq.~\eqref{eq:Sbare}. Although our final result for the
flow equation, and therefore the fixed points, is equivalent to the result obtained by Bergman and Halperin \cite{Bergman76}, our derivation emphasizes the close analogy to the field rescaling of the Ising field. More importantly, our derivation reveals that the renormalized Ising fixed point R and the spherical fixed point S are both characterized by a finite anomalous dimension $y_{\ast} < 0$ of the strain field, so that $Y_l$ diverges at these fixed points. 
In Sec.~\ref{sec:free_energy} we have shown that this has observable consequences for the elastic equation of state (i.e., the stress-strain relation) at the fixed points R and S: while to leading order in the strain the stress-strain relation is still linear (Hooke's law), the anomalous dimension of the strain field generates subleading non-analytic corrections to Hooke's law.

The methods developed in this work can be extended in several directions.
First of all, by using non-perturbative truncation schemes of the formally exact Wetterich equation based on the gradient expansion \cite{Berges02,Dupuis21} it should be possible to obtain more reliable predictions on the fixed-point structure in three dimensions. It is also possible to formulate the FRG procedure such that the external strain (and not the stress) is fixed during the RG flow. In this case we should adjust the scale-dependent homogeneous strain $E_{\Lambda}$ such that the one-point vertex in the vertex expansion of the
the average effective action $\Gamma_{\Lambda} [ \bar{\phi}, E_{\Lambda} + \bar{\epsilon} ]$  vanishes identically \cite{Schuetz06}. 
Finally, it should be
straightforward to apply our FRG approach to Euclidean quantum field theories
describing the effect of elastic fluctuations on quantum criticality~\cite{Zacharias15,Zacharias15b,Chandra20,Samanta22,Sarkar23}.

\section*{ACKNOWLEDGEMENTS}
	
We thank Andreas Rückriegel for discussions and the Deutsche Forschungsgemeinschaft (DFG, German Research Foundation) for financial support via  TRR 288-422213477.

\begin{appendix}

\section*{APPENDIX A: Effective action for volume strain}
\setcounter{equation}{0}
\renewcommand{\theequation}{A\arabic{equation}}

In this section, we will derive the effective action $S [ \phi, E ]$
for an Ising field $\phi$ coupled to the volume strain
$E ( \bd{r} ) = \sum_i E_{ ii} ( \bd{r} )$  given in 
Eq.~(\ref{eq:Sbare}) starting from the general coupled Ising-elastic action 
$S_{\phi}+ S_E  + S_{\phi , E }$ defined in Eqs.~(\ref{eq:Sphi}, \ref{eq:SE}, \ref{eq:allgemeines_3point_coupling}). As in Eq.~(\ref{eq:Eee}) we decompose
 $E_{ij} ( \bd{r} ) =
e_{ij} + \epsilon_{ij} ( \bd{r} )$ and write our bare action as
\begin{eqnarray}
	S [ \phi , \epsilon_{ij}, e_{ij} ] & = & \int d^D r \left[
	  \frac{c_0}{2} ( {\mathbf{\nabla}} \phi)^2 +
	\frac{r_0}{2}   \phi^2 +
	\frac{ u_0 }{ 4!} \phi^4 \right]
	\nonumber
	\\ 
	& + &   \frac{1}{2}\int d^D r  \sum_{ijkl}  \epsilon_{ij}     C_0^{ij,kl} \epsilon_{kl}  
	+  \frac{g_0}{2}  \int d^D r \phi^2  \epsilon 
	\nonumber
	\\
	& + &   \frac{\mathcal{V}}{2}   \sum_{ijkl} e_{ij}     C_0^{ij,kl} e_{kl}  
	+  \frac{g_0}{2} e  \int d^D r \phi^2  ,
\end{eqnarray}
where we have introduced the homogeneous and fluctuating parts of the volume strain,
\begin{subequations}
	\begin{align}
	 e & = \sum_i e_{ii} , \\
 \epsilon (\bd{r} ) & = \sum_i \epsilon_{ii}(\bd{r}).
	\end{align}
\end{subequations}
First, we consider the phonon contribution to the effective strain action. 
Using Eq.~\eqref{eq:relation_e_und_u} relating the finite-momentum part 
$\epsilon_{ij} ( \bd{r} )$  of the strain tensor to the
phonon displacement field $ \bd{u} ( \bd{r} )$ we may write
\begin{align}
& \frac{1}{2}   \int d^D r \sum_{ijkl} \epsilon_{ij}     C_0^{ij,kl} \epsilon_{kl} 
	=  \frac{1}{2} \int d^D r \sum_{ijkl} \partial_i u_j     C_0^{ij,kl} 
	\partial_k u_l.
\end{align}
Given the fact that the Ising field in 
Eq.~(\ref{eq:allgemeines_3point_coupling}) couples only to the volume strain
 \begin{equation}
E ( \bd{r} ) = \sum_i e_{ii} + \sum_i \epsilon_{ii} ( \bd{r} ) =
e + \mathbf{\nabla} \cdot \bd{u} ( \bd{r} ),
 \end{equation}
it is convenient to work with the effective 
action of this component of the strain  by
 performing a  constrained  integration over the phonon displacement field $\bd{u} ( \bd{r} ) $ for fixed
$\epsilon ( \bd{r} ) = \mathbf{\nabla} \cdot \bd{u} ( \bd{r} )$ and
$e = \sum_i e_{ii}$. For the finite-momentum phonon part the constraint of constant $\epsilon ( \bd{r} )$ can be enforced via 
an auxiliary integration,
 \begin{equation}
 \prod_{\bd{r}} 
\delta \Bigl( \epsilon (\bd{r}) - {\mathbf{\nabla}} \cdot \bd{u} ( \bd{r} ) \Bigr)
 = \int {\cal{D}} [ \lambda ]
 e^{ i \int d^D r  \lambda ( \bd{r} )
 [ \epsilon ( \bd{r} ) - \mathbf{\nabla} \cdot \mathbf{u}( \bd{r} ) ] },
 \end{equation}
where $\prod_{\bd{r}}$ denotes a product of
Dirac $\delta$-functions for each point of the suitably regularized space.
The Gaussian effective action $S^{\rm eff}_{\epsilon }$ of the volume strain field
$\epsilon ( \bd{r} )$ can then be written as
\begin{widetext}
	\begin{gather}
		e^{-S^{\rm eff}_{\epsilon}}\equiv\int {\cal{D}} [ \lambda ]
		e^{ i \int d^D r  \lambda ( \bd{r} ) 
			\epsilon ( \bd{r})  }	\int_{}\mathcal{D}[\bd{u}]e^{ -\frac{1}{2}\int d^D r  \sum_{ijkl} \partial_i u_j(\bd{r})     C_0^{ij,kl} \partial_{k}u_j(\bd{r})-i\int d^D r\lambda ( \bd{r} ) 
			\mathbf{\nabla} \cdot  \bd{u} ( \bd{r} )  }.
	\end{gather}
\end{widetext}
We now exchange the order of integrations and first perform the  Gaussian integration over the phonon displacements $\bd{u} ( \bd{r} )$. Therefore, we transform all fields to momentum space,
\phantom{a}\\
\phantom{a}
\vspace{-0.6cm}
\begin{subequations}
\begin{align}
		\bd{u} ( \bd{r} ) & =   \int_{\bd{k}\neq 0} e^{ i \bd{k} \cdot \bd{r} } \bd{u}_{\bd{k}} ,  
		\\
	\epsilon_{ij} ( \bd{r} ) &  =    \int_{\bd{k} \neq0} e^{ i \bd{k} \cdot \bd{r} } \epsilon_{ij, \bd{k}} , 
	\\
	\lambda ( \bd{r} ) & =   \int_{\bd{k}\neq 0} e^{ i \bd{k} \cdot \bd{r} } \lambda_{\bd{k}} ,  
\end{align}
 \end{subequations}
where $ \int_{ \bd{k} \neq 0 } =  \frac{1}{\cal{V}} \sum_{\bd{k} \neq 0}$. The Gaussian integration over the phonon field 
gives the usual Debye-Waller factor. Defining
 \begin{eqnarray}
 M^{ij} ( {\bd{k}} )  & = & \sum_{\alpha \beta} C_0^{ i \alpha, j \beta} k_{\alpha} k_{\beta}, \\
 v^i_{ - \bd{k} } & =  & \lambda_{ - \bd{k}}  k_{i},
 \label{eq:Mdef}
 \end{eqnarray}
we obtain
\begin{eqnarray}
 e^{- S^{\rm eff}_{\epsilon}  }  & =  & 
 \int {\cal{D}} [ \lambda ] e^{ i \int_{\bd{k} \neq 0} \lambda_{ - \bd{k}} \epsilon_{\bd{k}}  }
 \nonumber
 \\
 & \times &
  \int {\cal{D}} [ \bd{u} ]   e^{ -  \int_{\bd{k} \neq 0}[ \frac{1}{2} \sum_{ij} u_{i, - \bd{k} } M^{ij}( {\bd{k}} ) 
 u_{j, \bd{k} }  + \sum_i   v^i_{- \bd{k} } u_{ i , \bd{k}}   ]   }
 \nonumber
 \\
 & = &   \int {\cal{D}} [ \lambda ] 
 e^{  \int_{\bd{k} \neq 0}  [ 
  \frac{1}{2} \sum_{ij}  v^i_{ - \bd{k}}  M_{ ij } ( \bd{k}  )  v^j_{\bd{k}}   +  i \lambda_{ - \bd{k}} \epsilon_{\bd{k}}  ] },
 \label{eq:Seps2}
 \hspace{7mm}
 \end{eqnarray}
where $M_{ ij} (  \bd{k})$ is the  inverse of the matrix formed by the $M^{ij} ( {\bd{k}} )$,
i.e.,
 \begin{equation}
 \sum_j M_{ i j } ( \bd{k} ) M^{j k } ( \bd{k} ) = \delta_{ i}^{\; k},
 \end{equation}
which is the usual relation between the covariant and contravariant elements of a metric tensor.
Finally, we integrate over the auxiliary field $\lambda_{\bd{k}}$ and obtain
\begin{equation}
 S^{\rm eff}_{\epsilon} = \frac{\rho_0}{2 } \int_{\bd{k} \neq 0 }  \epsilon_{ - \bd{k}} \epsilon_{\bd{k}} .
 \end{equation}
where the  elastic stiffness  $\rho_0 $   is defined via  the contraction
 \begin{equation}
 \frac{1}{\rho_0 } =  \sum_{ij} M_{ ij } (  \bd{k} )  
  k_{i}  k_j .
  \label{eq:m0def}
 \end{equation}
Note that  $\rho_0$ 
has units of density (inverse volume) and 
can be identified with the
bare stiffness of finite-momentum elastic fluctuations. 
Keeping in mind that in momentum space
$ \epsilon_{\bd{k}} = i \bd{k} \cdot \bd{u}_{\bd{k}} $ the stiffness $\rho_0$ is proportional to the square of the 
 longitudinal sound velocity.
For an elastically isotropic system we may express the matrix $M^{ij} ( \bd{k} )$
in terms of the two Lam\'{e} parameters $\lambda$ and $\mu$ introduced Eq.~\eqref{eq:Cijk_Lame}, 
\begin{equation}
	M^{ij} ( \bd{k} )  = \mu \bd{k}^2\delta^{ij} +\left(\lambda+\mu\right)k^{i}k^{j},
\end{equation}
implying
\begin{equation}
	\rho_0=\lambda+2\mu\equiv K_0+\frac{4}{3}\mu,
\end{equation}
with bulk modulus
\begin{equation}
	K_0\equiv\lambda+\frac{2}{3}\mu.
\end{equation}

Finally, let us also calculate the contribution from fluctuations of the homogeneous (volume) strain $e = \sum_i e_{ii}$ to the effective action, which can be written as
\begin{align}
 & e^{-S^{\rm eff}_e}  = \int\mathcal{D}[e_{ij}] 
  \delta \bigl( e - \sum_{i} e_{ii} \bigr)
 e^{-\frac{\mathcal{V}}{2}  \sum_{ijkl} e_{ij}     C_0^{ij,kl} e_{kl} }
  \nonumber
	\\
	&  = \int_{}^{}\mathcal{D}[A]e^{iA e}\int\mathcal{D}[e_{ij}]e^{-\frac{\mathcal{V}}{2}   \sum_{ijkl} e_{ij}     C_0^{ij,kl} e_{kl} 
	 -iA    \sum_i e_{ii}}.
	 \label{eq:Seffe}
\end{align}
Expressing the contraction of the strain tensor in terms of
the Lam\'{e} parameters,
\begin{align}
	& \frac{\mathcal{V}}{2} \sum_{ijkl} e_{ij}C^{ij,kl}_0e_{kl}
	\nonumber
	\\
	= & 
	\frac{\lambda}{2}\sum_{i\neq j}^{}e_{ii}e_{jj}+\left(\frac{\lambda}{2}+\mu\right)\sum_{i}^{}e_{ii}^2+\mu \sum_{ i \neq j}e_{ij}^2,
\end{align}
and performing the Gaussian integrations in Eq.~(\ref{eq:Seffe})
we finally obtain
\begin{equation}
	S^{\rm eff}[e]=\mathcal{V}\frac{K_0}{2}e^2.
\end{equation}
Collecting all contributions, we eventually obtain the effective action of the Ising field, which is coupled to volume strain,
\begin{align} 
	 S [ \phi , \epsilon, e ]  & = \int d^D r \left[  \frac{c_0}{2} ( {\mathbf{\nabla}} \phi)^2 + 
	\frac{r_0}{2}   \phi^2 +\frac{ u_0 }{ 4!} \phi^4  \right]
	\nonumber \\
	&
	 +   \frac{1}{2} \int_{\bd{k}\neq 0}    \rho_0 \epsilon_{ - \bd{k}} \epsilon_{\bd{k}} +\mathcal{V}\frac{K_0}{2}e^2
	 + \frac{g_0}{2} e\int_{\kv}^{}\phi_{-\kv}\phi_{\kv}
	\nonumber
	\\
	& 
	+ \frac{g_0}{2} \int_{ \bd{k}_1} \int_{ \bd{k}_2} \int_{\bd{k}_3\neq 0}
	{\cal{V}} \delta_{ \bd{k}_1 + \bd{k}_2 + \bd{k}_3 ,0} \phi_{ \bd{k}_1 } \phi_{ \bd{k}_2 } \epsilon_{\bd{k}_3 },	
	\label{eq:S_full_appendix}
\end{align}
which is equivalent to Eq.~(\ref{eq:Sbare}) in Sec.~\ref{sec:Setup}.
Since the action \eqref{eq:S_full_appendix} is quadratic in the strain fields $\epsilon_{\bd{k}}$ and $e$, we may explicitly carry out the integration $\epsilon$ and $e$ to obtain the effective action of the Ising field
\begin{align}
	S_{\text{eff}}[\phi] & =\frac{1}{2}\int_{\kv}\left[r_0+c_0k^2\right]\phi_{ - \bd{k}}\phi_{ \bd{k}} \nonumber \\
	 & +\frac{1}{4!}\left[u_0-\frac{3 g_0^2}{\rho_{0}}\right]\int_{ \kv,\kv',\qv\neq 0}\phi_{ - \bd{k}}\phi_{ \bd{k}-\qv}\phi_{ - \bd{k}'}\phi_{ \bd{k}'+\qv}\nonumber\\
	& +\frac{1}{4!\mathcal{V}}\left[u_0-\frac{3g_0^2}{K_0}\right]\int_{ \bd{k}}\phi_{ - \bd{k}}\phi_{\kv}\int_{\kv'}^{}\phi_{ - \bd{k}'}\phi_{ \bd{k}'}.
	\label{eq:S_eff}
\end{align}
Note that at the  special value of the interaction given by
\begin{equation}
	u_0= \frac{3 g_0^2}{\rho_{0}}
\end{equation}
only the long-range interaction in the last line survives,
 \begin{align}
	S_{\text{eff}}[\phi]& =\frac{1}{2}\int_{\kv}\left[r_0+c_0k^2\right]\phi_{ - \bd{k}}\phi_{ \bd{k}} \nonumber \\
	& +\frac{u_0}{4!\mathcal{V}}\left[1-\frac{\rho_{0}}{K_0}\right]\int_{ \bd{k}}\phi_{ - \bd{k}}\phi_{\kv}\int_{\kv'}^{}\phi_{ - \bd{k}'}\phi_{ \bd{k}'}.\label{eq:S_eff_S}
\end{align}
However, the coefficient of the interaction is negative,
\begin{equation}
	1-\frac{\rho_{0}}{K_0}=1-\left(1+\frac{4}{3}\frac{\mu}{K_0}\right)=-\frac{4}{3}\frac{\mu}{K_0}<0,
\end{equation} 
so that the resulting free energy does not exist, because the functional integral diverges. 


\section*{APPENDIX B:  Truncation including all three-point and four-point vertices}

\setcounter{equation}{0}
\renewcommand{\theequation}{B\arabic{equation}}

In this appendix, we present the details of the FRG calculation discussed
in Sec.~\ref{sec:FRG3}, where we include all three-point and four-point vertices allowed by symmetry. The corresponding average effective action is given in Eq.~(\ref{eq:vertexp2}).
Neglecting again the momentum dependence of all vertices, we have to keep track of the flow of the two self-energies $\Sigma_{\Lambda} (0)$ and
$\Pi_{\Lambda} (0)$ and five coupling constants $g_{\Lambda}, u_{\Lambda},
h_{\Lambda}, v_{\Lambda}, w_{\Lambda}$ defined in Eqs.~(\ref{eq:gu}) and (\ref{eq:uvw}). 
Recall that $h_{\Lambda}$ is the three-point vertex of the strain field,
$v_{\Lambda}$ is the four-point vertex of the strain field, and $w_{\Lambda}$ is the mixed four-point vertex with two Ising and two strain legs. Although these couplings vanish in our bare action (\ref{eq:Sbare}), they are generated by the RG flow. 
\subsection{Flow equations}
\label{subsec:modflow}

If we fix the homogeneous strain, the flow of the conjugate stress $\sigma_{\Lambda}$ is determined by the flow equation shown in Fig.~\ref{fig:flow_one_point}.
In  Sec.~\ref{sec:FRG4} we have dropped the contributions from all vertices which vanish at the initial scale.
Here we take these vertices into account, but still neglect their momentum dependence. In this approximation, the scale-dependent stress satisfies
 \begin{equation}\label{eq:flow3stress}
 	\partial_{\Lambda}\sigma_{\Lambda}=\frac{g_{\Lambda}}{2}\int_{\kv}\dot{G}_{\Lambda}(\kv)+\frac{h_\Lambda}{2}\int_{ \bd{k}}\dot{F}_{\Lambda}(\kv).
 \end{equation}
If we set $h_{\Lambda} =0$ we recover the flow equation (\ref{eq:stresssharp})
of Sec.~\ref{sec:FRG4}.

Graphical representations of the exact flow equations for the Ising self-energy $\Sigma_{\Lambda} ( \bd{k} )$ and
the strain self-energy $\Pi_{\Lambda} ( \bd{k} )$ have already been 
shown in  Fig.~\ref{fig:flow_two_point} of  Sec.~\ref{sec:FRG4}.
If we neglect again the momentum dependence of all vertices,
these graphs represent the following  flow equations,
\begin{align}
 \partial_{\Lambda} \Sigma_{\Lambda} (0) & =  \frac{u_{\Lambda}}{2} \int_{\bd{q}} \dot{G}_{\Lambda} ( \bd{q} )
+ \frac{w_{\Lambda}}{2} \int_{\bd{q}} \dot{F}_{\Lambda} ( \bd{q} )
 \nonumber
 \\
 & - g_{\Lambda}^2 \int_{\bd{q}} [ {G}_{\Lambda} ( \bd{q} ) F_{\Lambda} ( \bd{q} )  ]^{\bullet},
 \label{eq:sigmaflowall}
 \\
 \partial_{\Lambda} \Pi_{\Lambda} (0) & = 
\frac{{v}_{\Lambda}}{2} \int_{\bd{q}} \dot{F}_{\Lambda} ( \bd{q} )
+ \frac{w_{\Lambda}}{2} \int_{\bd{q}} \dot{G}_{\Lambda} ( \bd{q} )
 \nonumber
 \\
& 
- g_{\Lambda}^2 \int_{\bd{q}} \dot{G}_{\Lambda} ( \bd{q} ) G_{\Lambda} ( \bd{q} ) \nonumber \\
&- h_{\Lambda}^2 \int_{\bd{q}} \dot{F}_{\Lambda} ( \bd{q} ) F_{\Lambda} ( \bd{q} ).
\label{eq:Piflow_full}
 \end{align}
If we set $h_{\Lambda} = v_{\Lambda} = w_{\Lambda} =0$ we recover
Eqs.~(\ref{eq:Sigma0sharp}) and (\ref{eq:Pi0sharp}) of Sec.~\ref{sec:FRG4}.

\begin{figure}[t]
 	\begin{center}
 		\centering
 		\includegraphics[width=0.45\textwidth]{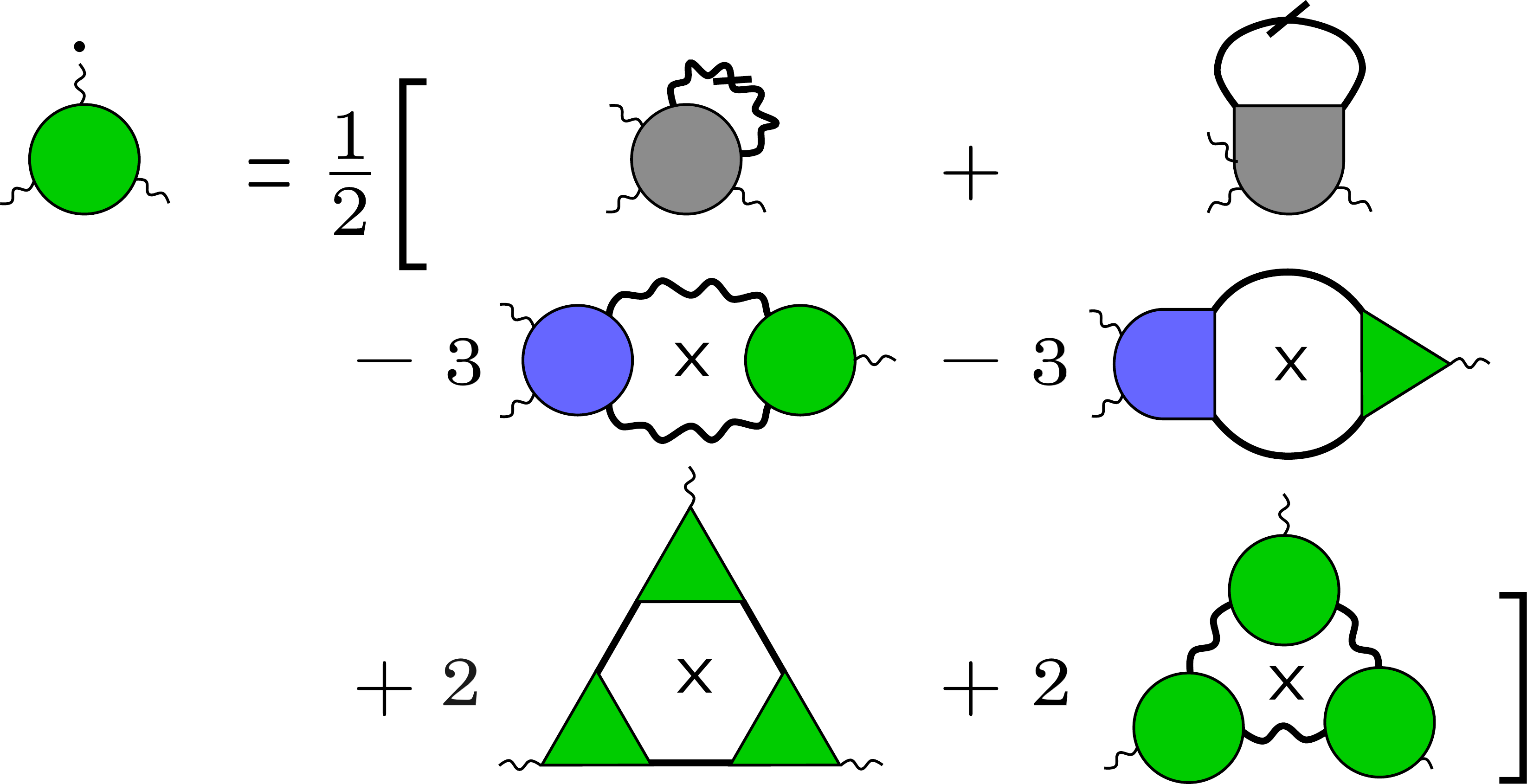}
 	\end{center}
 	\caption{%
 		Diagrammatic representation of the exact flow equation for the
 		strain three-point vertex $\Gamma^{\epsilon\epsilon\epsilon}_{\Lambda} (
 		\bd{k}_1 , \bd{k}_2 , \bd{k}_3 )$ which is represented by a green circle with
 		three external wavy legs.  The other symbols are explained
 		in the captions of Figs.~\ref{fig:flow_one_point}, \ref{fig:flow_two_point}, and
 		\ref{fig:flow_three_point}. The gray symbols represent various types of five-point vertices, which we neglect in our truncation.
 	}
 	\label{fig:flow_h_three_point}
 \end{figure}

Next, consider the flow of the three-point vertices $g_{\Lambda}$ and $h_{\Lambda}$.
A graphical representation of the exact flow equation for the
mixed three-point vertex $\Gamma^{\phi \phi \epsilon}_{\Lambda}$  is shown in
Fig.~\ref{fig:flow_three_point}, while the flow equation for the
strain three-point vertex $\Gamma^{\epsilon \epsilon \epsilon}_{\Lambda}$ 
is shown in Fig.~\ref{fig:flow_h_three_point}. 
 These flow equations (including all combinatorial factors and the correct signs) can be derived in a straightforward manner
from the general expressions for FRG flow equations given in Ref.~[\onlinecite{Kopietz10}].
 Setting the external momenta equal to zero, neglecting vertices with more than four external legs, and identifying $g_{\Lambda} = \Gamma_{\Lambda}^{\phi \phi \epsilon} (0,0,0)$ and $h_{\Lambda} =
 \Gamma^{\epsilon \epsilon \epsilon}_{\Lambda} (0,0,0)$ we obtain
  \begin{align}
 \partial_{\Lambda} g_{\Lambda} & = - u_{\Lambda} g_{\Lambda} 
 \int_{\bd{q}} \dot{G}_{\Lambda} ( \bd{q} ) G_{\Lambda} ( \bd{q} )
 - w_{\Lambda} h_{\Lambda} 
 \int_{\bd{q}} \dot{F}_{\Lambda} ( \bd{q} ) F_{\Lambda} ( \bd{q} )
 \nonumber
 \\
 &
 - 2 w_{\Lambda} g_{\Lambda}   \int_{\bd{q}} 
[ {G}_{\Lambda} ( \bd{q} ) F_{\Lambda} ( \bd{q} )  ]^{\bullet}
 + g_{\Lambda}^3 \int_{\bd{q}} [ {G}^2_{\Lambda} ( \bd{q} )  
 F_{\Lambda} ( \bd{q} ) ]^{\bullet} \nonumber \\
 &+ g_{\Lambda}^2 h_{\Lambda} 
\int_{\bd{q}} [  {F}^2_{\Lambda} ( \bd{q} )  
 G_{\Lambda} ( \bd{q} )   ]^{\bullet},
 \label{eq:gammaflow2}
 \\
 \partial_{\Lambda} h_{\Lambda} & = - 3 {v}_{\Lambda} h_{\Lambda} 
 \int_{\bd{q}} \dot{F}_{\Lambda} ( \bd{q} ) F_{\Lambda} ( \bd{q} )
  - 3 {w}_{\Lambda} g_{\Lambda} 
 \int_{\bd{q}} \dot{G}_{\Lambda} ( \bd{q} ) G_{\Lambda} ( \bd{q} )
 \nonumber
 \\
 & + 3 g^3_{\Lambda} \int_{\bd{q}} \dot{G}_{\Lambda} ( \bd{q} ) G^2_{\Lambda} ( \bd{q} )
+ 3 h_{\Lambda}^3  \int_{\bd{q}} \dot{F}_{\Lambda} ( \bd{q} ) F^2_{\Lambda} ( \bd{q} ). \label{eq:flow3strain}
 \end{align}
Recall that according to the product rule introduced in Eq.~(\ref{eq:productrule}) the symbol $[ {G}^n_{\Lambda} ( \bd{q} ) F_{\Lambda}^m ( \bd{q} )  ]^{\bullet}$ denotes the following combination of propagators and single-scale propagators,
 \begin{align}
[ {G}^n_{\Lambda} ( \bd{q} ) F_{\Lambda}^m ( \bd{q} )  ]^{\bullet} & = n \dot{G}_{\Lambda} ( \bd{q} )
 G_{\Lambda}^{n-1} ( \bd{q} ) F_{\Lambda}^m ( \bd{q} )
 \nonumber
 \\
 & +  m \dot{G}^n_{\Lambda} ( \bd{q} )
 \dot{F}_{\Lambda} ( \bd{q} ) F_{\Lambda}^{m-1} ( \bd{q} ).
 \end{align}

 \begin{figure}[t]
 	\begin{center}
 		\centering
 		\includegraphics[width=0.45\textwidth]{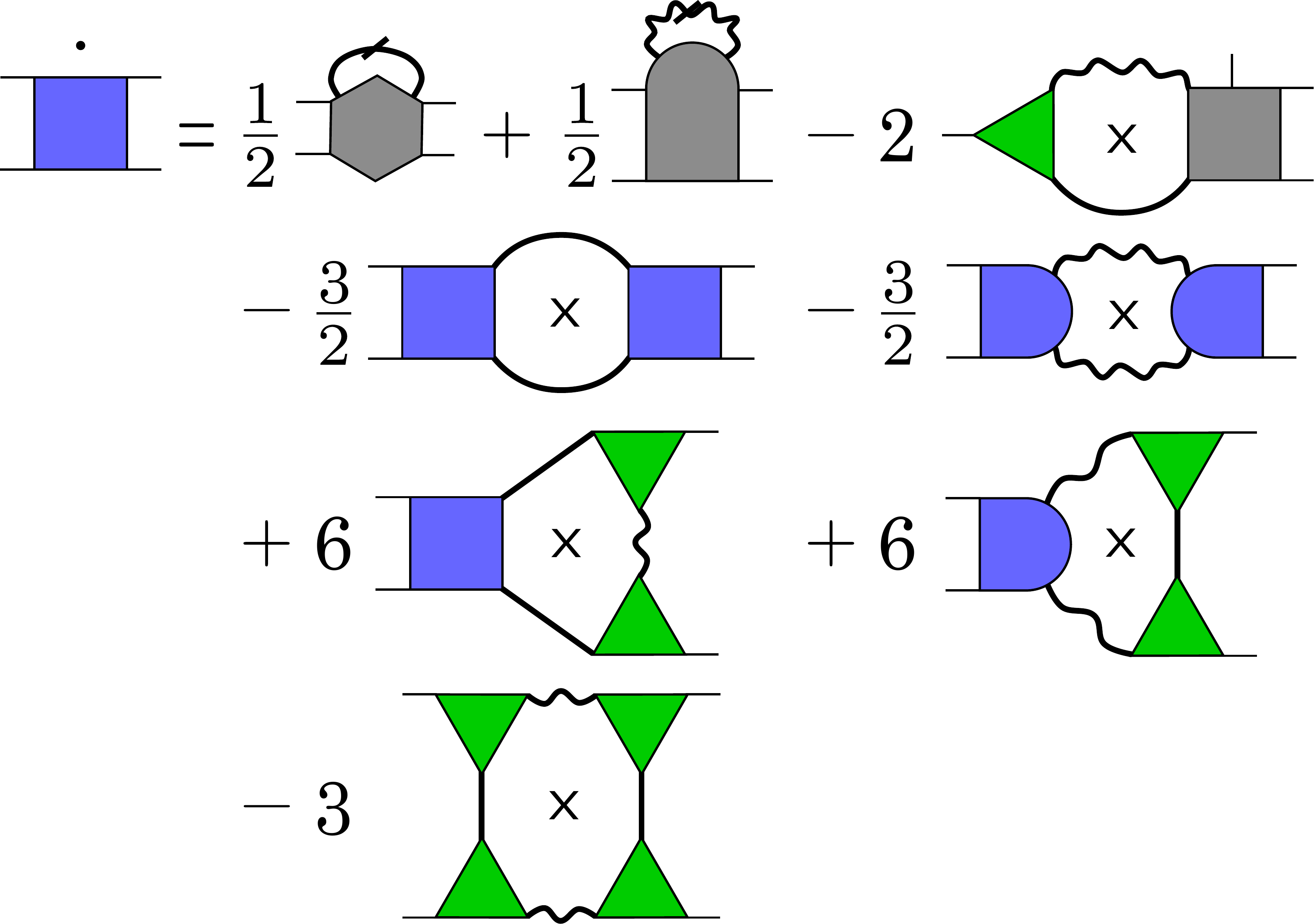}
 	\end{center}
 	\caption{%
 		Diagrammatic representation of the FRG flow equation for the
 		Ising four-point vertex $u_{\Lambda}= \Gamma^{\phi\phi\phi\phi}_{\Lambda} (
 		0,0,0,0)$ for vanishing external momenta.}
 	
 	\label{fig:flow_u_four_point}
 \end{figure}

Next, let us give the flow equation of the Ising four-point vertex.
Within the simplest truncation where the vertices which vanish at the initial scale are neglected the flow equation for $\Gamma^{\phi \phi \phi \phi}_{\Lambda} ( \bd{k}_1 ,
\bd{k}_2 , \bd{k}_3 , \bd{k}_4 )$ is given in Eq.~(\ref{eq:flow4}) and shown graphically in 
Fig.~\ref{fig:flow_four_point}.
\begin{figure}[t]
	\begin{center}
		\centering
		\includegraphics[width=0.45\textwidth]{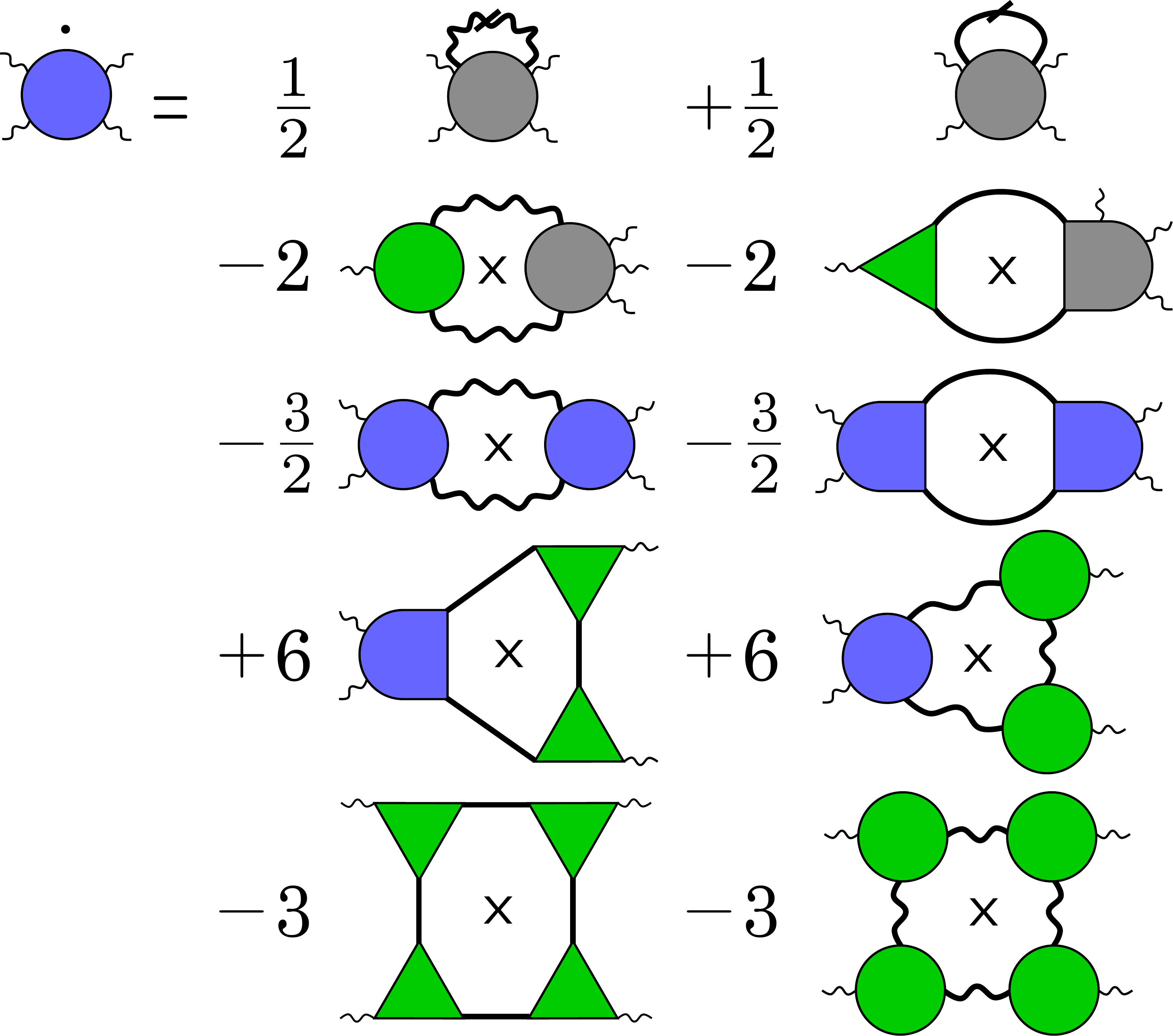}
	\end{center}
	\caption{%
		Diagrammatic representation of the exact FRG flow equation for the
		strain four-point vertex $ v_\Lambda = 
		\Gamma^{\epsilon\epsilon\epsilon\epsilon}_{\Lambda} (
		0,0,0,0)$ for vanishing external momenta.}
	\label{fig:flow_v_four_point}
\end{figure}
Note that for vanishing external momenta all permutations of the external legs indicated in Eq.~(\ref{eq:flow4}) give the same contribution to the flow equation. Since here we are only interested in the flow of the vertices for vanishing external momenta, we can simplify the graphical representation of the flow equations by multiplying the various contributions by the combinatorial factors generated by the permutations of the external legs. The improved flow equation for the Ising four-point vertex
 $u_{\Lambda} = \Gamma^{\phi \phi \phi \phi}_{\Lambda} ( 0,0,0,0 )$
is then given by
 \begin{align}
 & \partial_{\Lambda} u_{\Lambda}  = - 3 u_{\Lambda}^2  \int_{\bd{q}} \dot{G}_{\Lambda} ( \bd{q} ) G_{\Lambda} ( \bd{q} )
 - 3 w_{\Lambda}^2  \int_{\bd{q}} \dot{F}_{\Lambda} ( \bd{q} ) F_{\Lambda} ( \bd{q} )
 \nonumber
 \\ &
 + 6 u_{\Lambda} g_{\Lambda}^2  \int_{\bd{q}} [ G^2_{\Lambda} ( \bd{q} ) 
 F_{\Lambda} ( \bd{q} )  ]^{\bullet}
+  6 w_{\Lambda} g_{\Lambda}^2  \int_{\bd{q}} [ G_{\Lambda} ( \bd{q} ) 
 F^2_{\Lambda} ( \bd{q} )  ]^{\bullet}
 \nonumber
 \\ &
  - 6 g_{\Lambda}^4  \int_{\bd{q}} [  G^2_{\Lambda} ( \bd{q} ) 
 F^2_{\Lambda} ( \bd{q} )   ]^{\bullet}.\label{eq:flowuorder}
 \end{align}
Note that for $w_\Lambda =0$ this flow equation reduces to Eq.~(\ref{eq:usharp}).
A graphical representation of Eq.~(\ref{eq:flowuorder}) is shown in Fig.~\ref{fig:flow_u_four_point}. 

Finally, we also need the FRG flow of the
momentum-independent parts of the strain four-point vertex $v_{\Lambda} = \Gamma_{\Lambda}^{\epsilon \epsilon \epsilon \epsilon}  (0,0,0,0)$ and the
mixed four-point vertex  
 $w_{\Lambda} = \Gamma_{\Lambda}^{\phi \phi \epsilon \epsilon}  (0,0,0,0)$.
Graphical representations of the corresponding exact flow equations are 
shown in Figs.~\ref{fig:flow_v_four_point} and \ref{fig:flow_w_four_point}.
Within our truncation where vertices with more than four external legs are neglected, 
the flow equations reduce to
 \begin{align}
 \partial_{\Lambda} v_{\Lambda} & = 
 - 3 {v}_{\Lambda}^2  \int_{\bd{q}} \dot{F}_{\Lambda} ( \bd{q} ) F_{\Lambda} ( \bd{q} )
 - 3 w_{\Lambda}^2  \int_{\bd{q}} \dot{G}_{\Lambda} ( \bd{q} ) G_{\Lambda} ( \bd{q} )
 \nonumber
 \\ & 
 + 6 {v}_{\Lambda} h_{\Lambda}^2  \int_{\bd{q}} [ F^3_{\Lambda} ( \bd{q} )  ]^{\bullet}
 + 6 w_{\Lambda} g_{\Lambda}^2  \int_{\bd{q}} [ G^3_{\Lambda} ( \bd{q} )  ]^{\bullet}
 \nonumber
 \\
 & 
  - 6 h_{\Lambda}^4  \int_{\bd{q}} [  F^4_{\Lambda} ( \bd{q} ) ]^{\bullet}
 - 6 g_{\Lambda}^4  \int_{\bd{q}} [  G^4_{\Lambda} ( \bd{q} ) ]^{\bullet} ,
 \label{eq:flowustrain}
\end{align}

\newpage
\begin{widetext}
	\onecolumngrid
	\begin{figure}[t]
		\begin{center}
			\centering
			\includegraphics[width=0.85\textwidth]{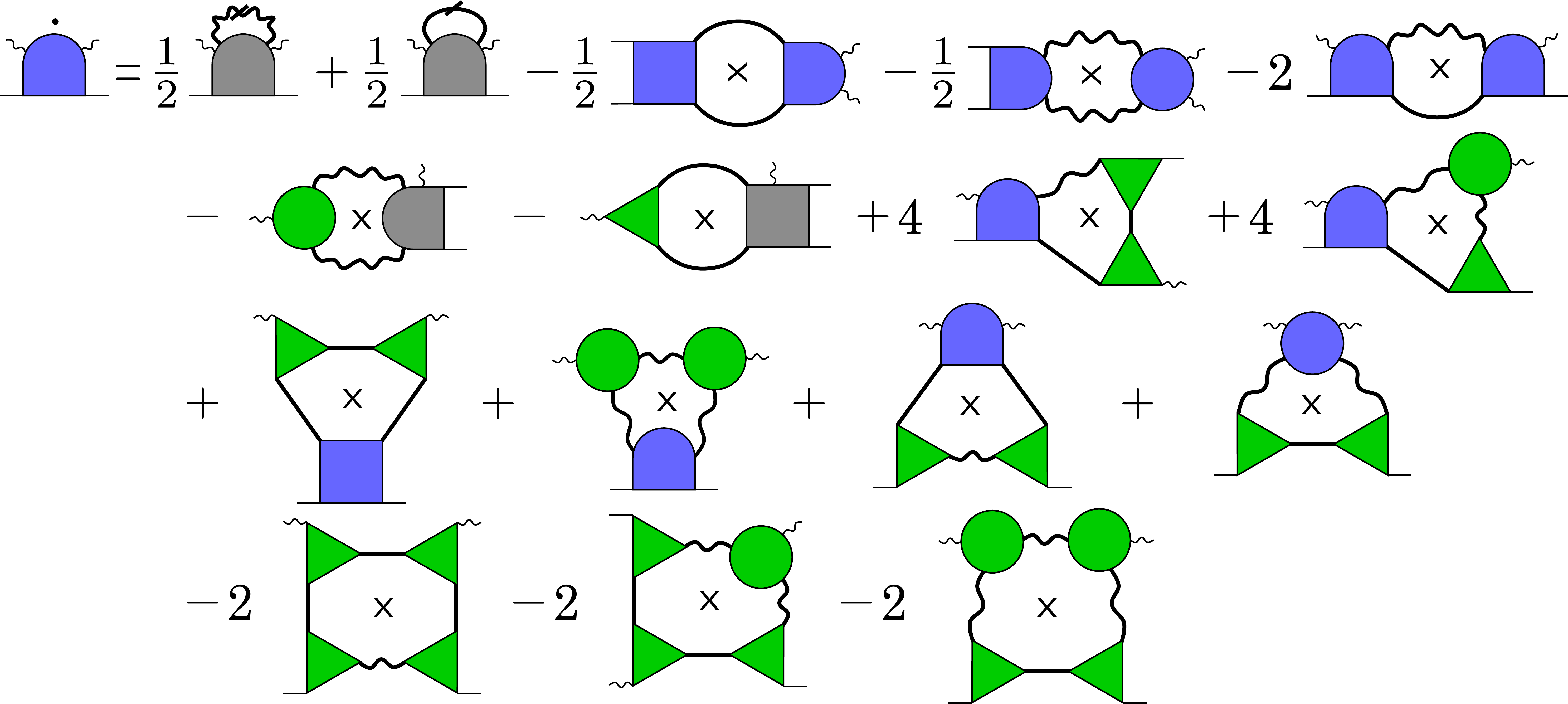}
		\end{center}
		\caption{%
			Diagrammatic representation of the exact FRG flow equation for the mixed four-point vertex $w_{\Lambda} = \Gamma_{\Lambda}^{\phi \phi \epsilon \epsilon}   ( 0,0,0,0 )$ for vanishing external momenta.}
		\label{fig:flow_w_four_point}
	\end{figure}
	\begin{align}
		\partial_{\Lambda} w_{\Lambda} & = - w_{\Lambda} u_{\Lambda} 
		\int_{\bd{q}} \dot{G}_{\Lambda} ( \bd{q} ) G_{\Lambda} ( \bd{q} )
		- w_{\Lambda} {v}_{\Lambda} 
		\int_{\bd{q}} \dot{F}_{\Lambda} ( \bd{q} ) F_{\Lambda} ( \bd{q} )
		- 2 w_{\Lambda}^2 \int_{\bd{q}} [ G_{\Lambda} ( \bd{q} ) F_{\Lambda} ( \bd{q} ) ]^{\bullet}  + u_{\Lambda} g_{\Lambda}^2  \int_{\bd{q}} [ G^3_{\Lambda} ( \bd{q} )    ]^{\bullet}
		\nonumber
		\\
		& + w_{\Lambda} h_{\Lambda}^2  \int_{\bd{q}} [ F^3_{\Lambda} ( \bd{q} )   ]^{\bullet}
		 + {v}_{\Lambda} g_{\Lambda}^2 \int_{\bd{q}}
		[ G_{\Lambda} ( \bd{q} ) F^2_{\Lambda} ( \bd{q} ) ]^{\bullet}
		+ {w}_{\Lambda} g_{\Lambda}^2 \int_{\bd{q}}
		[ G^2_{\Lambda} ( \bd{q} ) F_{\Lambda} ( \bd{q} ) ]^{\bullet} 
		+ 4 {w}_{\Lambda} g_{\Lambda}^2 \int_{\bd{q}}
		[ G^2_{\Lambda} ( \bd{q} ) F_{\Lambda} ( \bd{q} ) ]^{\bullet}
		\nonumber
		\\
		&
		+ 4 {w}_{\Lambda} g_{\Lambda} h_{\Lambda}  \int_{\bd{q}}
		[ G_{\Lambda} ( \bd{q} ) F^2_{\Lambda} ( \bd{q} ) ]^{\bullet} - 2 g_{\Lambda}^4  \int_{\bd{q}}
		[ G^3_{\Lambda} ( \bd{q} ) F_{\Lambda} ( \bd{q} ) ]^{\bullet}
		- 2 g_{\Lambda}^3 h_{\Lambda}
		\int_{\bd{q}}
		[ G^2_{\Lambda} ( \bd{q} ) F^2_{\Lambda} ( \bd{q} ) ]^{\bullet}
		- 2 g_{\Lambda}^2 h^2_{\Lambda}  \int_{\bd{q}}
		[ G_{\Lambda} ( \bd{q} ) F^3_{\Lambda} ( \bd{q} ) ]^{\bullet}. 
		\label{eq:floww}
	\end{align}
\end{widetext}

\twocolumngrid

\subsection{Flow of rescaled couplings}

To find the fixed points of the above system of flow equations
we should properly rescale the couplings. 
Apart from the rescaled couplings $r_l$ and $u_l$ previously defined in Eqs.~(\ref{eq:rldef}) and (\ref{eq:uldef}), we introduce
 \begin{subequations}
  \label{eq:rescaled2}
  \begin{align}
  {g}_l  & =  \frac{Z_l}{c_0}\sqrt{ \frac{ K_D \Lambda^{D-4} }{\rho_{\Lambda} }} g_\Lambda,\\
  {h}_l  & = \sqrt{ \frac{ K_D     \Lambda^D}{\rho^3_{\Lambda} } } h_{\Lambda}  ,\\
  {v}_l & =\frac{ K_D     \Lambda^D}{\rho^2_{\Lambda}} v_{\Lambda},
\\
 {w}_l  & 
= \frac{ K_D Z_{l}     \Lambda^{D-2} }{
   c_{0}  \rho_{\Lambda} }{w}_{\Lambda},
  \end{align}
  \end{subequations}
Neglecting again the momentum dependence of the self-energies, we may set
$Z_l \approx 1$ and hence $\eta_l =0$. We nevertheless retain $\eta_l$ in the flow equations below to indicate how they are modified for finite $\eta_l$. The modified flow equation for $r_l$ is then
 \begin{align}\label{eq:r_3d_flow}
  \partial_l r_l   = & ( 2 - \eta_l )  r_l + \frac{1}{2} \biggl[ \frac{ u_l}{ 1 + r_l  } +
   {w}_l\biggr]  - \frac{ {g}^2_l}{ 1 + r_l}.
 \end{align}
Using the flow Eq.~\eqref{eq:Piflow_full} for  the elastic self-energy $\Pi_{\Lambda} (0)$
we obtain the anomalous dimension of the strain field
 \begin{align}\label{eq:flowing_y_d3}
  y_l  &  = - \partial_l \ln Y_{\Lambda}   = \frac{1}{2}
  \left[ {v}_l   + {w}_l- \frac{ {g}_l^2}{ ( 1 + r_l )^2 }
   - {h}_l^2 \right].
   \end{align}
The rescaled three-point vertices satisfy the flow equations
\begin{align}
 \partial_l {g}_l & =   \frac{1}{2} \left( 4 - D  - 2 \eta_l - y_l \right) {g}_l 
 - \frac{1}{2} \left[ \frac{ u_l {g}_l }{ (1 + r_l )^2 } + {w}_l 
   {h}_l  \right]
  \nonumber
  \\
  &-  \frac{ 2  {w}_l {g}_l}{1 + r_l  } 
  + \frac{ {g}_l^3}{( 1 + r_l )^2 }
+ \frac{ {g}_l^2  {h}_l }{  1 + r_l } ,
\\
 \partial_l  {h}_l & =  - \frac{1}{2} \left( D +3 y_l \right)  {h}_l
 -  \frac{3}{2} \left[  {v}_l {h}_l
 + \frac{  {w}_l {g}_l }{ ( 1 + r_l )^2 } \right]
 \nonumber
 \\ 
 &  + \frac{ {g}_l^3}{( 1 + r_l )^3 } + {h}_l^3.
\end{align}  
Finally, the flow of the three distinct rescaled four-point vertices is given by
 \begin{align}
 \partial_l u_l & = ( 4 -D - 2 \eta_l ) u_l 
 - \frac{3}{2} \left[ \frac{ u_l^2}{( 1 + r_l )^2 }
 +{w}_l^2 \right]
 \nonumber
 \\
 &  +   6 \left[ \frac{  u_l {g}_l^2}{( 1 + r_l )^2 } 
  + \frac{{w}_l {g}_l^2}{  1+ r_l }
  \right] -  6 \frac{  {g}_l^4}{ ( 1 + r_l )^2 },
 \\
   \partial_l {v}_l & = - (  D + 2 y_l ) {v}_l 
   - \frac{3}{2} \left[ {v}_l^2 + \frac{ {w}_l^2}{ ( 1 + r_l )^2} \right]
 \nonumber
 \\
 &  +   6 \left[{v}_l  {h}_l^2+  \frac{{w}_l {g}_l^2}{ ( 1+ r_l )^3 } -  {h}_l^4     -   \frac{  {g}_l^4}{( 1 + r_l )^4 }  
  \right],
\end{align}
and
\begin{align}
  \hspace{-2mm}\partial_l {w}_l & =  ( 2  -D - \eta_l - y_l ) {w}_l 
  - \frac{1}{2} \left[ \frac{ {w}_l u_l}{( 1 + r_l )^2 } +  {w}_l  {v}_l
   \right] 
 \nonumber
 \\
 & -  \frac{ 2 {w}_l^2}{ 1 + r_l  }    + \frac{ u_l {g}_l^2}{ ( 1 + r_l )^3} + {w}_l  {h}_l^2 
 + 5 \frac{ {w}_l {g}_l^2}{ ( 1 + r_l )^2  }+\frac{ 4 {w}_l {g}_l {h}_l }{  1 + r_l }  \nonumber
 \\
 &+ \frac{ {v}_l {g}_l^2}{  1 + r_l } -  \frac{ 2 {g}_l^2}{ 1 + r_l }
 \left[ \frac{ {g}_l^2}{( 1 + r_l )^2} +  {h}_l^2 +
  \frac{ {g}_l {h}_l }{ 1 + r_l  } \right].   
   \label{eq:flow3w}
 \end{align} 
%
Note that we  recover the flow equations derived in Sec.~\ref{sec:fixed4}
by setting   $0=\eta_l = {h}_l = {v}_l = {w}_l$.

\begin{figure}[t]
	\begin{center}
		\centering
		\includegraphics[width=0.42\textwidth]{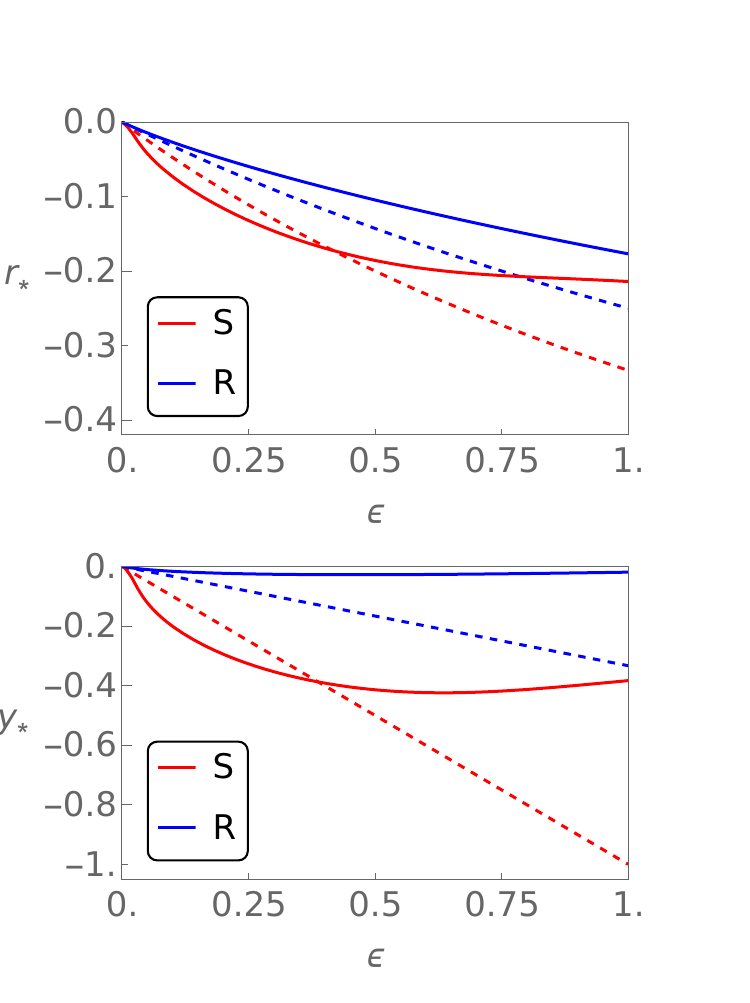}
	\end{center}
	\vspace{-6mm}
	\caption{Evolution of the fixed point values $r_{\ast}$ and $y_{\ast}$ at the renormalized Ising fixed point R (blue) and  the spherical fixed point S  (red)	traced from  $D=4$ ($\varepsilon =0$)  to $D=3$ ($\varepsilon = 1$). The dashed line represents the truncation used in Sec.~\ref{sec:FRG4}, while the solid lines represent the more sophisticated truncation including the couplings $h_l, v_l,$ and $w_l$ discussed in this appendix.
	}
	\label{fig:couplingsevol1}
\end{figure}

\subsection{Evolution of fixed points}
\label{sec:Fixedpoints_3_d}

\begin{figure}[t]
	\begin{center}
		\centering
		\includegraphics[width=0.42\textwidth]{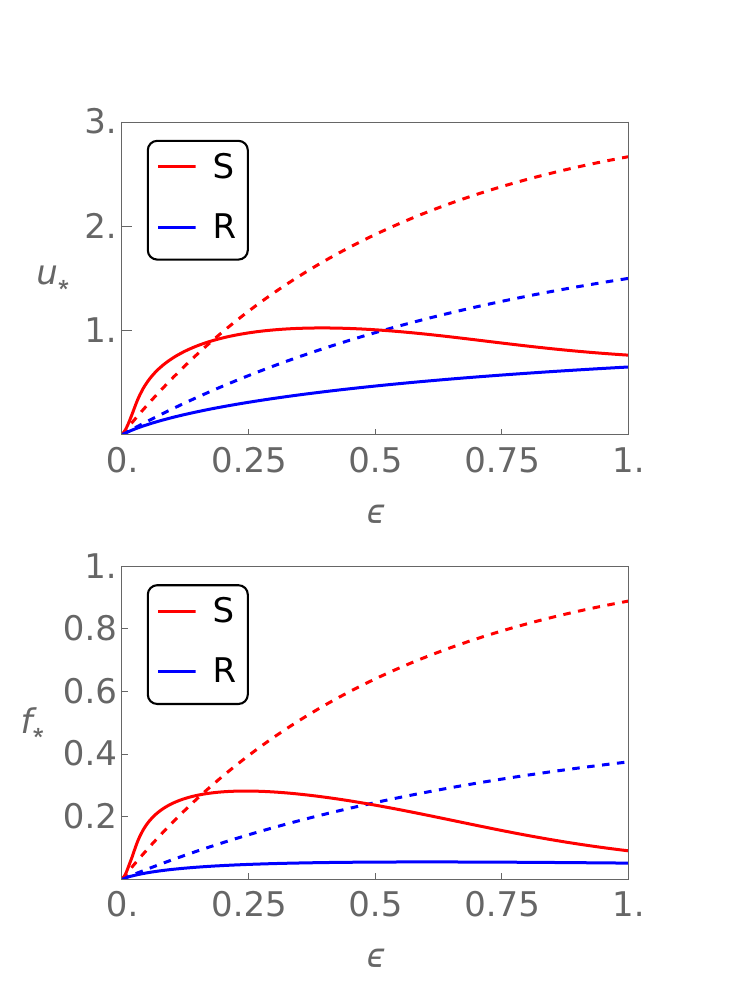}
	\end{center}
	\vspace{-6mm}
	\caption{Evolution of the fixed point values $u_{\ast}$ and $f_{\ast}\equiv g_{\ast}^2$ at the renormalized Ising fixed point R (blue) and  the spherical fixed point S  (red)	traced from  $D=4$ ($\varepsilon =0$)  to $D=3$ ($\varepsilon = 1$). The dashed line represents the truncation used in Sec.~\ref{sec:FRG4}, while the solid lines represent the more sophisticated truncation including the couplings $h_l, v_l,$ and $w_l$ discussed in this appendix.
	}
	\label{fig:couplingsevol2}
\end{figure}

By demanding that  the right-hand sides of the above flow equations for $r_l, g_l, h_l, u_l, v_l$, and $w_l$ vanish we obtain six coupled non-linear equations. Those equations have to be solved simultaneously to obtain the fixed points $r_{\ast}, g_{\ast}, h_{\ast}, u_\ast, v_{\ast},$ and $w_{\ast}$ of the RG flow within our truncation. 
The number of solutions of this system of equations depends on the parameter $\varepsilon = 4-D$. For small $\varepsilon $ we recover the four distinct fixed points G, I, R, and S discussed in Sec.~\ref{sec:fixed4}, since $h_{\ast}$, $v_{\ast}$, and $w_{\ast}$ are of higher order in $\varepsilon$. For $\varepsilon =1$, i.e. in three dimensions, we find numerically that the fixed point equations have additional solutions.  
The emergence of new fixed points of RG equations as a function of the dimensionality has also been observed in $O (N)$ models \cite{Yabunaka17}. 
An exhaustive classification of the solutions is beyond the scope of this work.
In fact, we believe that some of the fixed points are an unphysical artifact of our truncation.
Nonetheless, the four fixed points G, I, R, and S continue to exist in the entire interval
$0 < \varepsilon \leq 1$ so that we can continuously trace the evolution of these fixed points as a function of the system's dimensionality.
Because the Gaussian fixed point G and the Ising fixed point I remain unaffected by the inclusion of the couplings $h_l,v_l,w_l$, let us focus here on the evolution of the
renormalized Ising fixed point R and the spherical fixed point S. Our numerical results for the evolution of the fixed point values of
$r_{\ast}, g_{\ast}^2, u_{\ast}$ and $y_{\ast}$ 
as function of $\varepsilon = 4-D$  are shown on Fig.~\ref{fig:couplingsevol1} and Fig.~\ref{fig:couplingsevol2}. For comparison, the dashed lines represent the corresponding results obtained within the lowest order truncation used in Sec.~\ref{sec:FRG4}.

\begin{figure}[t]
	\vspace{-11mm}
	\begin{center}
		\centering
		\includegraphics[width=0.42\textwidth]{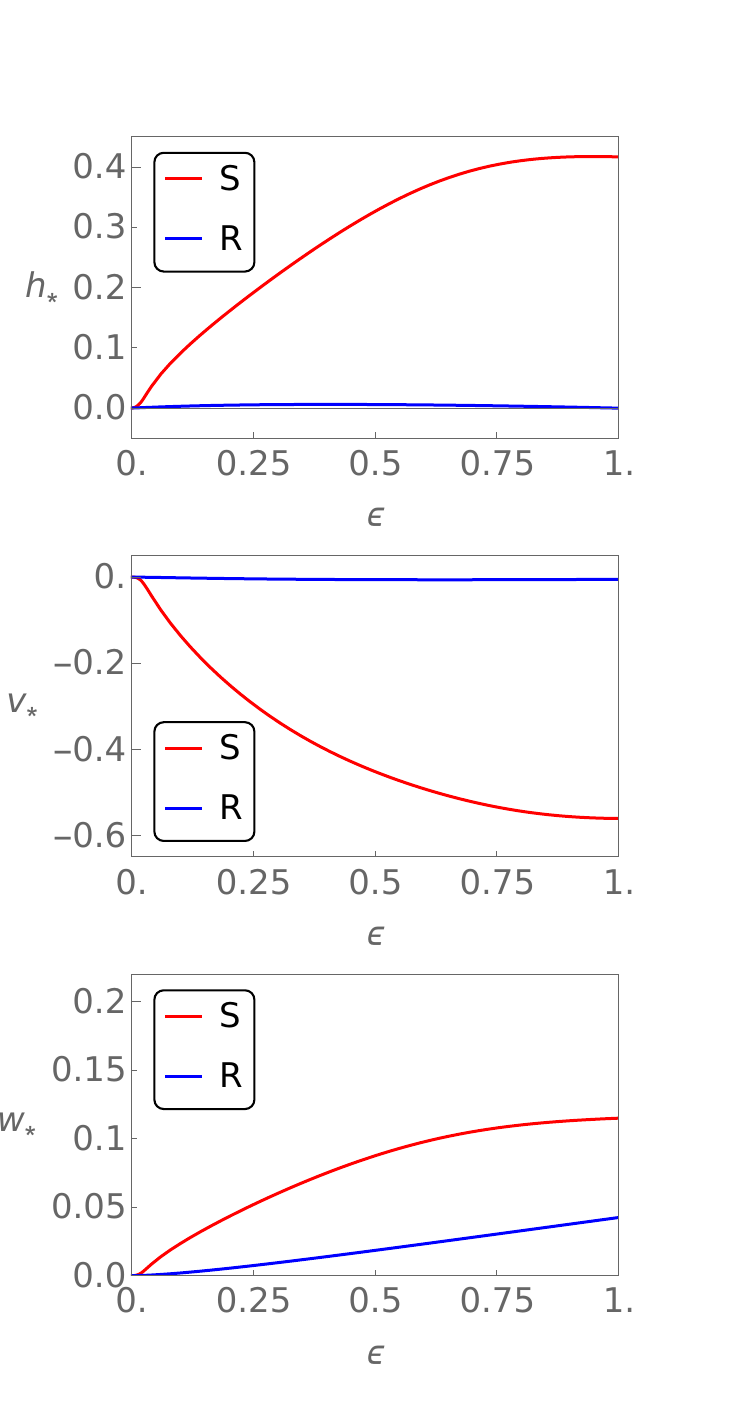}
	\end{center}
	 \vspace{-6mm}
	\caption{Fixed point values of $h_{\ast}$, $v_{\ast}$, $w_{\ast}$ traced form $\varepsilon =0$ (four dimensions)  to $\varepsilon = 1$
	(three dimensions).
	}
	\label{fig:hvw_up_to_3_dim}
\end{figure}

The evolution of the couplings $h_{\ast}, v_{\ast}$, and $w_{\ast}$, which are generated by the RG flow, is shown in 
Fig.~\ref{fig:hvw_up_to_3_dim}.
The limitations of the lowest order $\varepsilon$-expansion to obtain quantitative results in $D=3$ are obvious from these figures. Note also that for the spherical fixed point, the impact of the additional couplings is, in general, larger than for the renormalized Ising fixed point. Nevertheless, the fact that the fixed points found in Sec.~\ref{sec:FRG4} within a truncation which is controlled only to leading order in $\varepsilon$ continue to exist even for $\varepsilon =1$ indicates that the calculations of Sec.~\ref{sec:FRG4} can qualitatively describe the physics in three dimensions.

\subsection{Specific heat exponent}
\label{sec:Stabiliy_3_d}

\begin{figure}[b]
	\centering
	\includegraphics[width=0.475\textwidth]{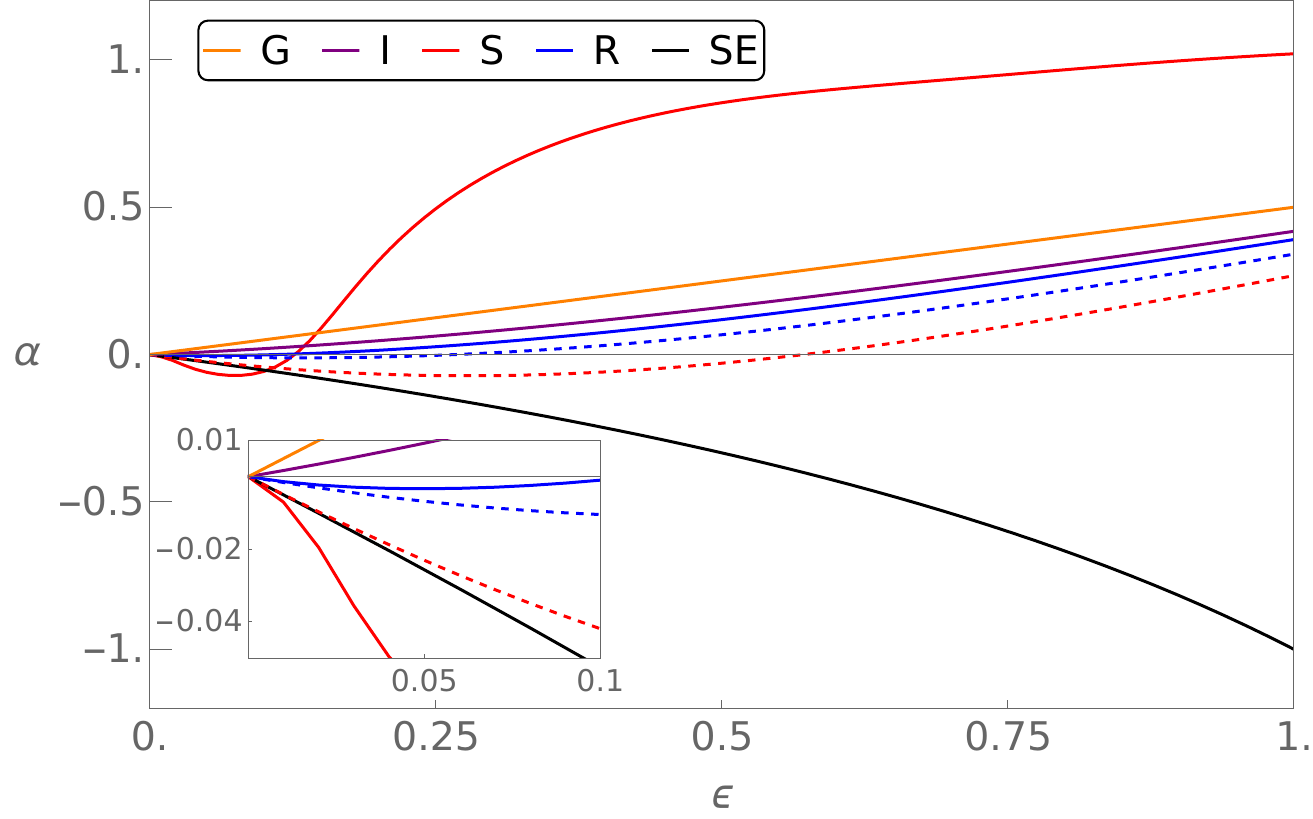}
	\caption{Plot of the specific heat exponent $\alpha$ as a function of $\varepsilon=4-D$ for each of the fixed points G, I, R, and S. The solid blue line depicts $\alpha_R$ within the truncation described in this appendix, while the the dashed blue line represents $\alpha_R$ using the simpler truncation of  Sec.~\ref{sec:FRG4}. Similarly, the solid red line depicts $\alpha_S$ using the truncation of this appendix, while the dashed red line represents $\alpha_S$ using the simpler truncation of Sec.~\ref{sec:FRG4}. The solid black line, labeled SE, shows the exact critical exponent of the spherical model given in Eq.~\eqref{eq:exact_spherical}. The solid orange line depicts $\alpha_G$ and the purple line $\alpha_I$, which are not modified by the inclusion of the additional couplings $h_l$, $v_l$,  and $w_l$. Therefore, the values of $\alpha_G$ and $\alpha_I$ can be obtained within the simple truncation of Sec.~\ref{sec:FRG4}.}
	\label{fig:alphaofepsilon}
\end{figure}

By linearizing the flow equations (\ref{eq:r_3d_flow}–\ref{eq:flow3w}) around the fixed points, we have determined the scaling variables and the corresponding eigenvalues.
We find that the scaling behavior close to G, I, and R  remains essentially the same as found in Sec.~\ref{sec:FRG4} because the number of relevant scaling directions is not affected by the additional couplings $h_l$, $v_l$, and $w_l$. On the other hand, the scaling in the
vicinity of the spherical fixed point S  undergoes a qualitative change
when we include the additional couplings
$h_l$, $v_l$ and $w_l$ acquiring two new relevant directions in $D=3$.
From the eigenvalue $\lambda_1 = 1 / \nu$ of the thermal scaling variable close to a given fixed point we obtain the specific heat exponent $\alpha = 2 - D \nu$. 
Our numerical results are summarized in Fig.~\ref{fig:alphaofepsilon}, where we show the evolution of $\alpha$ with dimensionality $\varepsilon=4-D$ for each of the fixed points G, I, R, and S.
From Fig.~\ref{fig:alphaofepsilon} we conclude that neither the approximation scheme of Sec.~\ref{sec:FRG4}, which is controlled to leading order  $\varepsilon$-expansion, nor the more elaborate truncation scheme described in this appendix 
are consistent with Fisher renormalization of the critical exponents beyond the leading order in $\varepsilon$. Specifically, since both $\alpha_G$ and $\alpha_I$ must be monotonically increasing functions of $\varepsilon$, Fisher rescaling predicts that $\alpha_S$ and $\alpha_{R}$ should be monotonically decreasing functions of $\varepsilon$, a behavior reflected in the exact specific heat exponent of the spherical model in Eq.~\eqref{eq:exact_spherical}. However, Fig.~\ref{fig:alphaofepsilon} shows that neither $\alpha_S$ nor $\alpha_{R}$ follows this monotonic decrease, regardless of the truncation used. This suggests that both truncation schemes fail to fully capture the qualitative behavior of the true physical critical exponents beyond the $\varepsilon$-expansion.

\hspace{0.2mm}
\end{appendix}


\begin{thebibliography}{99}
\bibitem{footnoteel}
For a system with cubic symmetry, the elastic tensor has three independent elements which are conventionally denoted by $C_{11} = C_0^{xx,xx} = C_0^{yy,yy} = C_0^{zz,zz}$,
$C_{12} = C_0^{xx,yy} = C_0^{yy,zz} = C_0^{zz,xx}$ and $C_{44} = C_0^{xy,xy} = C_0^{yz,yz} =
C_0^{zx, zx}$. For an isotropic system the two principal shear moduli are equal,
$\mu = C_{44} = ( C_{11} - C_{12} )/2$ and the bulk modulus can be written as
$K_0 = ( C_{11} + 2 C_{12} )/3$; the stiffness $\rho_0$ defined in Eq.~(\ref{eq:rhodef})
is then given by $\rho_0 = C_{11}$.
\end{thebibliography}
\end{document}